\documentclass[aps,pra,showpacs,preprintnumbers,amsmath,amssymb,twocolumn]{revtex4}
\usepackage{graphicx, color, graphpap}
\usepackage{amsmath}
\usepackage{enumitem}
\usepackage{amssymb}
\usepackage{amsthm}
\usepackage{pstricks}
\usepackage{float}
\usepackage{multirow}
\usepackage{hyperref}
\usepackage{subfigure}
\usepackage{caption2}
\long\def\ca#1\cb{} 

\newcommand{\tr}{{\mathsf{tr}}}

\renewcommand{\geq}{\geqslant}
\renewcommand{\leq}{\leqslant}

\newcommand{\be}{\begin{equation}}
\newcommand{\ee}{\end{equation}}
\newcommand{\ba}{\begin{array}}
\newcommand{\ea}{\end{array}}

\newtheoremstyle{example}{\topsep}{\topsep}%
{}
{}
{\bfseries}
{.}
{   }
{\thmname{#1}\thmnumber{ #2}}
\theoremstyle{example}

\theoremstyle{definition}

\begin{document}
\parskip=8pt
\title{Improved quantum entropic uncertainty relations}
\author{Zhihua Chen$^{1}$}
\author{Zhihao Ma$^{2}$}
\author{Yunlong Xiao$^{3}$}
\author{Shao-Ming Fei$^{4,5*}$}
\affiliation{
$^1$Department of Applied Mathematics, College of Science, Zhejiang
University of Technology, Hangzhou, 310014, China\\
$2$Department of Mathematics, Shanghai Jiaotong University, Shanghai 200240, China\\
$3$Department of Mathematics and Statistics and Institute for Quantum Science and Technology, University of Calgary, Calgary, Alberta T2N 1N4, Canada\\
$^4$School of Mathematical Sciences, Capital Normal University, Beijing 100048, China\\
$^5$Max-Planck-Institute for Mathematics in the Sciences, 04103 Leipzig, Germany}

\thanks{Corresponding author (email: *feishm@cnu.edu.cn)}


\begin{abstract}
We study entropic uncertainty relations by using stepwise linear functions and quadratic functions. Two kinds of improved uncertainty lower bounds are constructed: the state-independent one based on the lower bound of Shannon entropy and the tighter state-dependent one based on the majorization techniques. The analytical results for qubit and qutrit systems with two or three measurement settings are explicitly derived, with detailed examples showing that they outperform the existing bounds. The case with the presence of quantum memory is also investigated.
\end{abstract}
\maketitle


\section{introduction}

At the heart of quantum theory, uncertainty principle reveals the intrinsic difference between classical physics and quantum physics: experimentalists' (in)ability to perform precise measurements on a quantum system is fundamentally limited in the quantum world. For example, the Heisenberg uncertainty principle \cite{Heisenberg} tells us that the more information one can gain about the momentum of a quantum particle implies less certainty about its position and vice versa. This principle gives rise to wide applications in quantum cryptographic tasks, as well as in detection of entanglement,  Einstein-Podolsky-Rosen (EPR) steering, nonlocality and quantum metrology \cite{Guhne, Schneeloch, Giovanetti, Koashi, Renes, Tomamichel}.


The uncertainty principle concerns the uncertainty of a quantum variable. In terms of different uncertainty measures, one can formulate different uncertainty relations. Pioneered by Kennard \cite{Kennard1927} (see also the work of Weyl \cite{Weyl1927}), many physicists have employed variances to express uncertainty relations \cite{Robertson, Schrodinger1930, Pati, Xiao2016W, Xiao2016M, Xiao2016N, Xiao2017I}. Another approach to describe the uncertainty is to use differential entropy. In 1975, Bia\l{}ynicki-Birula and Mycielski \cite{Birula1975} obtained the first entropic formulation of uncertainty relation. Entropic uncertainty relations were later studied by means of shannon entropy for finite-dimensional quantum systems \cite{Deutsch}: Consider a quantum state $\rho$ belonging to an $N$-dimensional Hilbert space $\mathcal{H}_{N}$, and observables $\mathcal{A}_1$ and $\mathcal{A}_2$, the eigenstates $|a^{(1)}_{j}\rangle$ ($|a^{(2)}_{j}\rangle$) of $\mathcal{A}_1$ ($\mathcal{A}_2$) constitute an orthinormal basis in $\mathcal{H}_{N}$. According to Born's rule, the probability of measuring  $\mathcal{A}_1$ ($\mathcal{A}_2$) on $\rho$ with the $j$-th outcome is give by $p_j(\rho)=\tr(\rho |a^{(1)}_j\rangle\langle a^{(1)}_j|)$ [$q_j(\rho)=\tr(\rho |a^{(2)}_j\rangle\langle a^{2}_j|)$], and the corresponding Shannon entropy is defined by $H(\mathcal{A}_1)=-\sum\limits_{j}p_{j}(\rho)\log_2 p_{j}(\rho)$
[$H(\mathcal{A}_2)=-\sum\limits_{j} q_{j}(\rho)\log_2 q_{j}(\rho)$]. If the non-degenderate observables $\mathcal{A}_1$ and $\mathcal{A}_2$ do not have a common eigenstate, $H(\mathcal{A}_1)+H(\mathcal{A}_2)$ is bounded from below, and the bound depends only on the overlap between observables' eigenvectors, $e_{ij}=|\langle a^{(1)}_i|a^{(2)}_j\rangle|$. Denote $c=\max |\langle a^{(1)}_i|a^{(2)}_j\rangle|$. The  entropic uncertainty relation reads
\begin{align}
H(\mathcal{A}_1)+H(\mathcal{A}_2)\geq -2 \log_2\frac{1+c}{2}.
\end{align}

The lower bound has been improved by Maassen and Uffink \cite{Maassen}. They proved that, for any quantum state $\rho$, it holds
\begin{align}
H(\mathcal{A}_1)+H(\mathcal{A}_2)\geq -2\log_2 c.
\end{align}
Later the bound was further improved \cite{de},
\begin{equation}\label{deb}
H(\mathcal{A}_1)+H(\mathcal{A}_2)\geq
\left\{
\begin{aligned}
-2\log_2 c, & \hspace{0.5cm} \rm{if} \hspace{0.5cm} 0 < c \leq \frac{1}{\sqrt{2}}\\
H_1(c), & \hspace{0.5cm} \rm{if} \hspace{0.5cm} \frac{1}{\sqrt{2}}\leq c\leq c^*\\
F(c), & \hspace{0.5cm} \rm{if} \hspace{0.5cm} c^*\leq c\leq 1\\
\end{aligned}
\right.
\end{equation}
where $F(c)=-(1+c)\log_2 \frac{1+c}{2}-(1-c)\log_2 \frac{1-c}{2}$
and $H_1(c)=- P_A \log_2 P_A-(1-P_A)\log_2 (1-P_A)- P_B \log_2 P_B-(1-P_B)\log_2 (1-P_B),$
with $P_A=\cos^2\alpha,$ $P_B=\cos^2(\gamma-\alpha),$ $c=\cos\gamma,$
$\alpha$ is a numerical solution of the equation
\begin{equation}
\begin{aligned}
\nonumber
0=&\sin(2\alpha)\log_2\frac{1+\cos(2\alpha)}{1-\cos(2\alpha)}\\ \nonumber
+&\sin[2(\alpha-\gamma)]\log_2\frac{1+\cos(2(\alpha-\gamma))}{2(1-\cos^2(\alpha-\gamma))},
\end{aligned}
\end{equation}
such that $\alpha\neq \frac{\gamma}{2}$ and $\alpha\neq \frac{\gamma}{2}+\frac{\pi}{4}$, and $c^*$
is determined numerically in \cite{de}.

The uncertainty principle in the presence of quantum memory was later introduced by M. Berta {\it et. al.} \cite{Berta2010}, in which the measured system is correlated with another quantum system.
For any bipartite state $\rho_{AB}$, Bob's uncertainty about Alice's measurement outcomes of observables $\mathcal{A}_1$ and $\mathcal{A}_2$ on Alice's system $A$ is bounded,
\begin{align}
H(\mathcal{A}_1|B)+H(\mathcal{A}_2|B)\geq -2\log_2 c + H(A|B),
\end{align}
where $H(A|B) = H(\rho_{AB}) - H(\rho_{B})$ is the conditional entropy, $H(\mathcal{A}_1|B)$ stands for Bob's ignorance about the Alice measurement $\mathcal{A}_1$ on system $A$, given that Bob can access to the quantum memory $B$ [similarly for $H(\mathcal{A}_2|B)$].


However, in the context of the uncertainty principle, the measures of uncertainty should satisfy the following conditions \cite{Narasimhachar2016}: first, the uncertainty cannot decrease under randomly chosen symmetry transformations; second, the uncertainty cannot decrease under classical processing channels (followed by recovery). Friedland {\it et. al.} \cite{Friedland} therefore defined a measure of uncertainty from any non-negative Schur-concave functions including entropic functions such as Shannon entropy and R\'enyi entropy, and formulated the so-called ``universal uncertainty relations'' \cite{Friedland}.
Majorization technique was also used to construct such uncertainty relations \cite{Puchaa} as
in Ref. \cite{Friedland}. Meanwhile, many efforts have been made to improve the uncertainty relations \cite{Coles,Rudnicki,Li, de, Liu, Xiao, Wehner,Coles2,Yuan, Xiao2016QM}.

In this paper we improve the lower bounds for entropic uncertainty relations by polynomial functions. Besides improving the uncertainty relations, we provide the insight that the mutually unbiased bases can be used to form a conservation law. The uncertainty relations with quantum memory are also investigated.

\section{Setting up the stage}






Given two probability vectors $\mathbf{x}=\{x_1, x_2, \cdots, x_N\}$ and $\mathbf{y}=\{y_1, y_2, \cdots, y_N\}$, arranged in descending order, the vector $\mathbf{x}$ is said to be majorized by $\mathbf{y}$, $\mathbf{x}\prec \mathbf{y}$, if $\sum\limits_{i=1}^{i=k} x_i \leq \sum\limits_{i=1}^{i=k} y_i  ~(k=1, 2, \cdots, N-1)$ and $\sum\limits_{i=1}^N x_i=\sum\limits_{i=1}^N y_i.$
A function $\mathbf{f}: \mathbb{R}_{+}^N \rightarrow \mathbb{R}$ is said to be Schur concave if
$\mathbf{f}(\mathbf{x})\geq \mathbf{f}(\mathbf{y})$ whenever $\mathbf{x}\prec \mathbf{y}$.
Both Shannon entropy and R\'enyi entropy are all Schur concave functions.
It has been shown in Refs. \cite{Puchaa, Friedland} that
for two measurement probability distributions $\mathbf{p}=\{p_1(\rho), p_2(\rho), \cdots, p_N(\rho)\}$ and
$\mathbf{q}=\{q_1(\rho), q_2(\rho), \cdots, q_N(\rho)\},$ one has
$\mathbf{p}\otimes \mathbf{q} \prec \mathcal{\omega}$, which implies
$$\mathbf{f}(\mathbf{p}\otimes \mathbf{q})\geq \mathbf{f}(\mathcal{\omega}),$$
where $\mathcal{\omega}=\{\Omega_1, \Omega_2-\Omega_1, \cdots, \Omega_N-\Omega_{N-1},0,\cdots, 0\}$ is state-independent, $\mathbf{f}$ is any nonnegative Schur concave function. The term $\Omega_{k}=\max\limits_{\mathcal{I}_k}\max\limits_{\rho}\sum\limits_{(m,n)\in\mathcal{I}_k}p_m(\rho)q_n(\rho)$ with $\mathcal{I}_k \subset [N]\times [N]$ being a subset of $k$ distinct pairs of indices $(m,n)$ and $[N]$ the set of the natural numbers from 1 to $N$.

Note that, for pure state $|\psi\rangle$ and observables $\mathcal{A}_1$ and $\mathcal{A}_2$ with eigenstates $|a^{(1)}_j\rangle$ and $|a^{(2)}_j\rangle$ respectively, one has
$p_j(|\psi\rangle)=|\langle a^{(1)}_j|\psi\rangle|^2$ and $q_j(|\psi\rangle)=|\langle a^{(2)}_j|\psi\rangle|^2$. The uncertainty relation becomes
\begin{equation}\label{Ru282}
\min\limits_{|\psi\rangle}H(\mathcal{A}_1)+H(\mathcal{A}_2)\geq \tilde{H}(\tilde{\omega}),
\end{equation}
where $\tilde{\omega}=(s_1,s_2-s_1,\cdots,s_N-s_{N-1}),$ $\tilde{H}(\tilde{\omega})=-\sum\limits_{k=1}^N (s_k-s_{k-1})\log_2(s_k-s_{k-1}),$ $s_0=0,$ $s_k=\max\{\|\hat{U}^{(1,k)}\|, \|\hat{U}^{(2,k-1)}\|, \cdots, \|\hat{U}^{(k,1)}\|\},$
$\|\hat{U}^{(n,m)}\|=\max \{\|U(I,J)\|: I,J \subset \{1,2,\cdots, N\},|I|= n,|J|= m\}$ and
$\|U(I,J)\|$ is the operator norm: the maximal singular value of $U(I,J)=\{\langle a^{(1)}_i|a^{(2)}_j\rangle\}_{i\in I, j\in J}$.

For observable $\mathcal{A}_i$ with eigenstates $|a^{(i)}_j\rangle,$ we denote $U^{(i)}$ the matrix with the $j-th$ column given by $|a^{(i)}_j\rangle,$ i.e. $U^{(i)}=\{|a^{(i)}_j\rangle\}$.
Let $S_k$ be the maximal square of operator norms calculated for the rectangular matrices of size $d\times(k + 1)$, formed by $k + 1$ columns taken from the concatenation of all $L$ matrices $\{U^{(i)}\}_{i=1}^L$,
\begin{equation}\nonumber
S_k=\max\{\sigma_1^2(|a^{(i_1)}_{j_1}\rangle, |a^{(i_2)}_{j_2}\rangle, \cdots,|a^{(i_{k+1})}_{j_{k+1}}\rangle )\}
\end{equation}
where $\sigma_1^2$ is the square of operator norms, the maximum runs over all subsets ${(i_1,j_1),(i_2,j_2),...,(i_{k+1},j_{k+1})}$ of cardinality $k + 1$ of set $\{1,2,...,L\}\times \{1,2,...,d\}$. The following uncertainty relations hold:
\begin{equation}\label{Ru283}
\min\limits_{|\psi\rangle}\sum\limits_i H(\mathcal{A}_i)\geq \tilde{H}(\tilde{\omega}),
\end{equation}
where $\tilde{\omega}=\{1,S_1-1, S_2-S_1,\cdots, S_{dL}-S_{dL-1}\}.$

The strong entropic uncertainty relations for multiple measurements are given in Ref. \cite{Xiao}:
for any given $d-$ dimensional mixed quantum state $\rho$ and $N$ measurements $\mathcal{A}_k=\{ |a_j^{(k)}\rangle \}$ $(k=1,2,\cdots, N),$
\begin{equation}\label{Xiaob}
\begin{aligned}
\sum\limits_k H(\mathcal{A}_k)\geq H(\omega),
\end{aligned}
\end{equation}
where $\omega=(\Omega_1, \Omega_2-\Omega_1, \cdots, 1-\Omega_a)$ with $a$ being the smallest index such that $\Omega_{a+1}=1,$
and $\Omega_k=(\frac{S_k}{N})^N,$
$S_k=\max\limits_{\sum\limits_{x=1}^N s_x=k+N-1}\{\lambda_1[U(s_1,s_2,\cdots, s_N)]\},$
with $\lambda_1[U(s_1,s_2,\cdots, s_N)]$ being the maximal eigenvalue of
\begin{equation}
\begin{aligned}
\nonumber
U(s_1,s_2,\cdots, s_N)=
\left(
\begin{array}{cccc}
I_{s_1}& U_{12} & \cdots & U_{1N}\\
U_{21} & I_{s_2} & \cdots & U_{2N}\\
\vdots & \vdots & \vdots& \vdots\\
U_{N1} & U_{N2} &\cdots & I_{s_N}\\
\end{array}
\right).
\end{aligned}
\end{equation}
The matrices $U_{ij}$ are defined by the subsets
$\{|a^{(1)}_{i_1}\rangle,|a^{(1)}_{i_2}\rangle,\cdots, |a^{(1)}_{i_{s_1}}\rangle\}$, $ \{|a^{(2)}_{i_1}\rangle,|a^{(2)}_{i_2}\rangle,\cdots, |a^{(2)}_{i_{s_2}}\rangle\},$ $\cdots,$ $\{|a^{(N)}_{l_1}\rangle,|a^{(N)}_{l_2}\rangle,\cdots, |a^{(N)}_{l_{s_N}}\rangle\}$ with $s_1 + s_2 +\cdots + s_N=k+N-1$. For instance,
\begin{equation}
\begin{aligned}
\nonumber
U_{12}=\left(
\begin{array}{c}
\langle a^{(1)}_{i_1}\\
\vdots\\
\langle a^{(1)}_{i_{s_1}}\\
\end{array}
\right).
\left(
\begin{array}{cccc}
|a^{(2)}_{j_1}\rangle & |a^{(2)}_{j_2}\rangle & \cdots & |a^{(2)}_{j_{s_2}}\rangle\\
\end{array}
\right).
\end{aligned}
\end{equation}
$U_{13}, U_{14}, \cdots, U_{N-1,N}$ are constructed similarly.

For any given qubit state $\rho$ with spectral decomposition $\rho= p|r\rangle\langle r|+(1-p)|r_{\perp}\rangle\langle r_{\perp}|$, and two measurements $\mathcal{A}_1$ and $\mathcal{A}_2,$ an improved bound was given in Ref.\cite{Yuan},
\begin{equation}\label{Yuanb}
\begin{aligned}
H(\mathcal{A}_1)+H(\mathcal{A}_2)\geq H_s(\frac{\sqrt{2P-1}(2c-1)+1}{2})+S(\rho),
\end{aligned}
\end{equation}
where $P=2p^2-2p+1$ is the purity of the state.
The bound in Ref.\cite{Korzekwa} is
\begin{equation}\label{korze}
\begin{aligned}
H(\mathcal{A}_1)+H(\mathcal{A}_2)\geq -2\log_2 c + 2 S(\rho)[1+\log_2 c],
\end{aligned}
\end{equation}
In this work, we give a tighter uncertainty relation by using the eigenvalues of quantum states and the transition from one measurement basis to the other one for mixed states.
Also we give a tighter uncertainty relations based on the lower bound of Shannon entropy.


\section{Entropic uncertainty relation with state-independent bound}

We now investigate entropic uncertainty relations with bounds given by polynomial functions.
Let $H_s(x)=-x \log_2 x-(1-x)\log_2(1-x)$ and $H_s(x,y) = -x \log_2 x - y \log_2 y - (1- x - y)\log_2(1-x-y)$.
For $H_s(x)$ and stepwise linear function, we divide the interval $0\leq x\leq 1$ into $n$ equal parts,
in each part $\frac{i-1}{n}\leq x\leq \frac{i}{n},$  the linear function $\mathbf{P}^{(i)}_1(x)$ is determined by
the two points $(\frac{i-1}{n}, H_s(\frac{i-1}{n}))$ and  $(\frac{i}{n}, H_s(\frac{i}{n})).$
Actually the difference between $H_s(x)$ and the stepwise linear function $\mathbf{P}^{(i)}_1(x),~\frac{i-1}{n}\leq x\leq \frac{i}{n},~ 1\leq i\leq n$,
decreases as the number of the equal parts $n$ increases.
Similarly, for $H_s(x,y)$ and stepwise linear function, we divide the region $\{0\leq x\leq 1, 0\leq y\leq 1-x\}$ into $n^2$ equal triangle region: $\frac{i-1}{n}\leq x\leq \frac{i}{n},\frac{j-1}{n}\leq y\leq \frac{i+j-1}{n} - x$ or $\frac{i-1}{n}\leq x\leq \frac{i}{n},\frac{i+j-1}{n} - x\leq y\leq \frac{j}{n}, 1\leq i,j\leq n.$
In each triangle region, the linear function $P_1^{(i,j)}(x,y)$ is determined by three points, $(\frac{i-1}{n},\frac{j-1}{n}, H_s(\frac{i-1}{n},\frac{j-1}{n}))$, $(\frac{i}{n},\frac{j-1}{n}, H_s(\frac{i}{n},\frac{j-1}{n}))$ and $(\frac{i-1}{n},\frac{j}{n}, H_s(\frac{i-1}{n},\frac{j}{n}))$ or
$(\frac{i}{n},\frac{j}{n}, H_s(\frac{i}{n},\frac{j}{n}))$, $(\frac{i}{n},\frac{j-1}{n}, H_s(\frac{i}{n},\frac{j-1}{n}))$ and $(\frac{i-1}{n},\frac{j}{n}, H_s(\frac{i-1}{n},\frac{j}{n})).$ The difference between $H_s(x,y)$ and the stepwise linear function $P_1^{(i,j)}(x,y)$ also decreases as $n$ increases.

We have the following relations:
\begin{equation}
\begin{aligned}
\nonumber
& H_s(x) \geq \mathbf{P}_2(x),~ \text{or}\\ \nonumber
& H_s(x) \geq \mathbf{P}^{(i)}_1(x),~\frac{i-1}{n}\leq x\leq \frac{i}{n},~ 1\leq i\leq n \nonumber
\end{aligned}
\end{equation}
and
\begin{equation}
\begin{aligned}
\nonumber
&H_s(x,y) \geq \mathbf{P}_2(x,y),~ \text{or}\\ \nonumber
&H_s(x,y) \geq \mathbf{P}^{(i,j)}_1(x,y),~1\leq i\leq n,~ 1\leq j\leq n-i+1,
\end{aligned}
\end{equation}
where
\begin{equation}
\begin{aligned}
&\mathbf{P}_2(x)=2 [1 - x^2 - (1-x)^2],\\ \nonumber
&\mathbf{P}^{(i)}_1(x)=n[(x-\frac{i-1}{n})H_s(\frac{i}{n})-(x-\frac{i}{n})H_s(\frac{i-1}{n})],\\ \nonumber
&\mathbf{P}_2(x,y)=2 [1 - x^2 - y^2 - (1-x-y)^2]
\end{aligned}
\end{equation}
and
\begin{eqnarray}
\mathbf{P}^{(i,j)}_1(x,y) \nonumber
=\left\{
\begin{aligned}
&a_{ij} x + b_{ij} y + c_{ij},~ x\in h_i,~ y \in u_{ij}(x) , \\
&a'_{ij} x + b'_{ij} y + c'_{ij},~ x\in h_i,~ y \in v_{ij}(x),
\end{aligned}
\right.
\end{eqnarray}
where $h_i=(\frac{i-1}{n},\frac{i}{n}),$ $u_{ij}(x)=(\frac{j-1}{n}$, $\frac{i+j-1}{n} - x)$, $v_{ij}(x)=( \frac{i+j-1}{n} - x,\frac{j}{n})$,
and the coefficients $a_{ij}$, $a'_{ij},$ etc., are given by
\begin{align}
[a_{ij},b_{ij},c_{ij}]^T=\mathcal{D}^{-1}\mathcal{H}^T,\quad [a'_{ij},b'_{ij},c'_{ij}]^T=\mathcal{D}^{'-1}\mathcal{H}^{'T}, \notag
\end{align}
where
\begin{eqnarray}
\begin{aligned}
&\mathcal{D}=D[\frac{i-1}{n},\frac{j-1}{n},\frac{i}{n},\frac{j-1}{n},\frac{i-1}{n},\frac{j}{n}],\\ \nonumber
&\mathcal{D}'=D[\frac{i}{n},\frac{j-1}{n},\frac{i-1}{n},\frac{j}{n},\frac{i}{n},\frac{j}{n}],\\ \nonumber
&\mathcal{H}=[H_s(\frac{i-1}{n},\frac{j-1}{n}),H_s(\frac{i}{n},\frac{j-1}{n}),H_s(\frac{i-1}{n},\frac{j}{n})],\\ \nonumber
&\mathcal{H}'=[H_s(\frac{i}{n},\frac{j-1}{n}),H_s(\frac{i-1}{n},\frac{j}{n}),H_s(\frac{i}{n},\frac{j}{n})],\\ \nonumber
&D[x_1,y_1,x_2,y_2,x_3,y_3]=
\left(
\begin{array}{ccc}
x_1& y_1 & 1\\
x_2& y_2 & 1\\
x_3& y_3 & 1\\
\end{array}
\right).
\end{aligned}
\end{eqnarray}


It is also worth noticing that, when a quantum memory is  {\it in absentia}, entropic uncertainty relations are entirely specified by the overlap matrix $U$ defining the transition from one measurement basis to the another one, $U=(\langle a^{(1)}_i|a^{(2)}_j\rangle)_{i,j=1}^N$.
If we make the same unitary transformations to the two measurement bases simultaneously, i.e.
$|a^{(1)'}_j\rangle= \mathbf{T}|a^{(1)}_j\rangle$ and $|a^{(2)'}_j\rangle= \mathbf{T}|a^{(2)}_i\rangle$ for some basis transformation $T$, the uncertainty lower bound will not be changed.

\noindent{\textit{Theorem I:}} For any pure state $|\psi\rangle$ and observables given by the bases $\mathcal{A}_k=\{ |a_j^{(k)}\rangle \}$, $k=1,2,\cdots, N$, we have strengthened
uncertainty lower bounds by the quadratic function,
\begin{equation}\label{tm1}
\begin{aligned}
\sum\limits_k H(\mathcal{A}_k)
\geq &\min\limits_{|\psi\rangle} \sum\limits_{k=1}^N \mathbf{P}_2[p^{(k)}_i(|\psi\rangle)]\\
\geq &\min\limits_{|\psi\rangle}  \sum\limits_{k=1}^N 2[1-\sum\limits_i [p^{(k)}_i(|\psi\rangle)]^2]
\end{aligned}
\end{equation}
or by the stepwise linear function,
\begin{equation}\label{tm2}
\begin{aligned}
\sum\limits_k H(\mathcal{A}_k)
\geq &\min\limits_{|\psi\rangle,i} \sum\limits_{k=1}^N \mathbf{P}^{(i)}_1[ p^{(k)}_j(|\psi\rangle)],
\end{aligned}
\end{equation}
where $p^{(k)}_j(|\psi\rangle)=|\langle a_j^{(k)}|\psi\rangle|^2$, $k=1,2,\cdots, N.$

To show that (\ref{tm1}) and (\ref{tm2}) give better lower bounds of uncertainties, let us consider the following detailed cases.

\noindent{\textit{Case I:}} First consider the most simple case: quibt states $|\psi\rangle$ and two measurement $\mathcal{A}_k=\{ |a_j^{(k)}\rangle \}$,  $k=1,2$.
We have $\mathbf{p}=\{p_1,p_2\}$ and $\mathbf{q}=\{q_1,q_2\}$ with $p_j=|\langle \psi|a_j^{(1)}\rangle|^2$, $q_j=|\langle \psi|a_j^{(2)}\rangle|^2$.
Denote $\langle a_1^{(1)}|a_1^{(2)}\rangle=\cos\gamma$ and $|\langle \psi|a_1^{(1)}\rangle|=\cos\theta$.
From (\ref{tm1}) we obtain the following uncertainty relations:
\begin{equation}\label{tm10}
\begin{aligned}
&H(\mathcal{A}_1)+H(\mathcal{A}_2)\\
\geq & \min\limits_{\theta} \{\mathbf{P}_2[ p_j(|\psi\rangle)]+\mathbf{P}_2[q_j(|\psi\rangle)]\}\\
\geq & \min\limits_{\theta} (4\cos^2\theta\sin^2\theta + 4 \cos^2(\theta-\gamma)\sin^2(\theta-\gamma))\\
\geq & 1-|\cos 2\gamma|
\end{aligned}
\end{equation}
by quadratic function. As for the stepwise linear function, the interval $0 \leq \cos^2\theta \leq 1$ is divided into $n$ equal parts. Hence,
$\mathbf{P}_1^{(i)}(\cos^2\theta)=n[(\cos^2\theta-\frac{i-1}{n})l_i-(\cos^2\theta-\frac{i}{n})l_{i-1}]$ with
$l_{i-1}=H_s(\frac{i-1}{n})$ and  $l_i=H_s(\frac{i}{n})$. Correspondingly, $\cos^2(\theta-\gamma)$ is also divided into $n$
parts determined by the points $(t_{i-1},s_{i-1})$ and $(t_i, s_i)$ with $t_{i-1}=\cos^2(\theta_{i-1}-\gamma),$ $t_i=\cos^2(\theta_i-\gamma),$ $s_{i-1}=H_s[\cos^2(\theta_{i-1}-\gamma)],s_i=H_s[\cos^2(\theta_{i}-\gamma)],$ $\theta_i=\arccos\sqrt{\frac{i}{n}},$ $1\leq i\leq n$. Therefore, $\mathbf{P}_1^{(i)}[\cos^2(\theta-\gamma)]
=\frac{\cos^2(\theta-\gamma)-t_{i-1}}{t_i-t_{i-1}}s_i-\frac{\cos^2(\theta-\gamma)-t_i}{t_i-t_{i-1}}s_{i-1}$. From (\ref{tm2}) we have
\begin{equation}\label{tm20}
\begin{aligned}
&H(\mathcal{A}_1)+H(\mathcal{A}_2)\\
\geq & \min\limits_{\theta,i} \{\mathbf{P}_1^{(i)}[ p_1(|\psi\rangle)]+\mathbf{P}_1^{(i)}[ q_1(|\psi\rangle)]\}\\
\geq & \min\limits_{\theta,i} \{\mathbf{P}_1^{(i)}(\cos^2\theta)+\mathbf{P}_1^{(i)}[\cos^2(\theta-\gamma)]\} \\
\geq & \min\limits_i L_i,
\end{aligned}
\end{equation}
where $L_i$ are given by following.
Let $\theta^{*}_i$ be the extreme points of $\mathbf{P}_1^{(i)}(\cos^2\theta)+\mathbf{P}_1^{(i)}[\cos^2(\theta-\gamma)]$. We have
\begin{equation}
\begin{aligned}
&\tan2\theta^{*}_i=\frac{\sin 2\gamma(s_i-s_{i-1})}{\cos 2\gamma(s_i-s_{i-1})+n(l_i-l_{i-1})(t_i-t_{i-1})}.
\end{aligned}
\end{equation}
If $\frac{i-1}{n}\leq \cos^2(\theta^{*}_i)\leq \frac{i}{n}$, i.e. $\min\{\theta_{i-1},\theta_i\}\leq \theta_i^*\leq \max\{\theta_{i-1},\theta_i\},$ we have
\begin{equation}
\begin{aligned}
L_i=&\min\{\mathbf{P}_1^{(i)}(\cos^2\theta_i^{*})+\mathbf{P}_1^{(i)}[\cos^2(\theta_i^{*}-\gamma)],\\ \nonumber
&\mathbf{P}_1^{(i)}(\cos^2\theta_i)+\mathbf{P}_1^{(i)}[\cos^2(\theta_i-\gamma)],\\ \nonumber
&\mathbf{P}_1^{(i)}(\cos^2\theta_{i-1})+\mathbf{P}_1^{(i)}[\cos^2(\theta_{i-1}-\gamma)]\},
\end{aligned}
\end{equation}
otherwise
\begin{equation}
\begin{aligned}
L_i=&\min\{\mathbf{P}_1^{(i)}(\cos^2\theta_i)+\mathbf{P}_1^{(i)}[\cos^2(\theta_i-\gamma)],\\ \nonumber
&\mathbf{P}_1^{(i)}(\cos^2\theta_{i-1})+\mathbf{P}_1^{(i)}[\cos^2(\theta_{i-1}-\gamma)]\}.
\end{aligned}
\end{equation}

Figure 1 shows our bounds with respect to $c$: bound (\ref{tm10}) by the quadratic function is represented by the solid line,and  bound (\ref{tm20}) by the stepwise linear function is represented by the medium size dotted line with $n=64$. The bound (\ref{Yuanb}) derived in Ref. \cite{Yuan} is plotted by the small-size dotted line, which is almost identical to the bound (dot-dashed line) obtained in Ref. {\color[rgb]{0,0,1}\cite{Rudnicki}}. Meanwhile, the result (\ref{deb}) obtained in Ref. \cite{de} is given by the  big size dotted line for $c^*\leq c\leq 1$, which is almost identical to the bound(thick solid line) using mathematica program, and the bound (\ref{Xiaob}) appeared in Ref. \cite{Xiao} is denoted by dashed line, which is just a little less than the bound (\ref{Yuanb}). The optimal bound is given by the thick solid line. From Fig. 1 we see that our bound by the quadratic function is better than the bounds in Ref.  \cite{Rudnicki}, the bound (\ref{Xiaob}) and (\ref{Yuanb}) for $c<0.79$. Our bound by the stepwise linear function, with $n=64$, is better than the bounds in Ref.  \cite{Rudnicki}, the bound (\ref{Xiaob}) and (\ref{Yuanb}) for $\frac{1}{\sqrt{2}}\leq c\leq 1.$ It is a good approximation to the exact value of $H(\mathcal{A}_1)+H(\mathcal{A}_2)$.

\begin{figure}[htb]
\centering
\includegraphics[width=5.5cm]{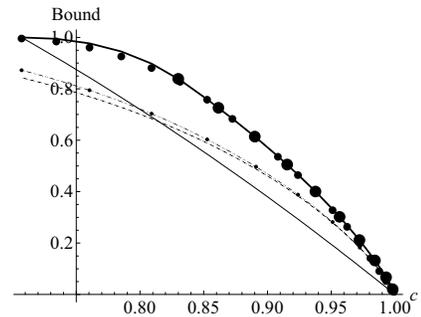}
\caption{The solid line is for our bound by the quadratic function; the medium size dotted line for our bound by the stepwise linear function; the small size dotted line for the the bound (\ref{Yuanb}), the dot-dashed line for the bound given in Ref. \cite{Rudnicki}; the big size dotted line for (\ref{deb}), the dashed line for the bound given in (\ref{Xiaob}) and the thick solid line for $H(\mathcal{A}_1)+H(\mathcal{A}_2)$ using mathematica program.}
\end{figure}

\noindent{\textit{Case II:}} Qubit state with three measurements $\mathcal{A}_k$ given by $\{ |a_j^{(k)}\rangle \}~ (k=1,2,3)$. Denote
$\mathbf{p}=\{p_1,p_2\},$ $\mathbf{q}=\{q_1,q_2\}$ and $\mathbf{r}=\{r_1,r_2\} $ with $p_j=|\langle \psi|a_j^{(1)}\rangle|,$ $q_j=|\langle \psi|a_j^{(2)}\rangle|$ and $r_j=|\langle \psi|a_j^{(3)}\rangle|$.
Set $q=\rm{arccos}|\langle a^{(1)}_1|a_1^{(2)}\rangle|$, $r=\rm{arccos}|\langle a^{(1)}_1|a_1^{(3)}\rangle|$, $g=\rm{arg}\langle a^{(1)}_1|a_1^{(2)}\rangle-\rm{arg}\langle a^{(1)}_2|a_1^{(2)}\rangle$ and
$h=\rm{arg}\langle a^{(1)}_1|a_1^{(3)}\rangle-\rm{arg}\langle a^{(1)}_2|a_1^{(3)}\rangle$. From (\ref{tm1}) we obtain
\begin{eqnarray}\label{iia}
\begin{aligned}
&H(\mathcal{A}_1)+H(\mathcal{A}_2)+H(\mathcal{A}_3)\\
=&-p_1\log_2 p_1-p_2\log_2 p_2-q_1\log_2 q_1-q_2\log_2 q_2\\
&-r_1\log_2 r_1-r_2\log_2 r_2\\
\geq &\min\limits_{\theta,\alpha} (\mathbf{P}_2(p_1)+\mathbf{P}_2(q_1)+\mathbf{P}_2(r_1))\\
\geq & \min\{\mathbf{P}_2(p_1)+\mathbf{P}_2(q_1)+\mathbf{P}_2(r_1)|_{\theta=\frac{m_1\pi}{2},\alpha=\frac{m_2\pi}{2}},\\
&\mathbf{P}_2(p_1)+\mathbf{P}_2(q_1)+\mathbf{P}_2(r_1)|_{\theta=\theta^*,\alpha=\alpha^*}\}
\end{aligned}
\end{eqnarray}
with
$$
\tan2\alpha^*=\frac{\sin^2q\cos^2q\sin2g+\sin^2r\cos^2r\sin2h}{\sin^2q\cos^2q\cos2g+\sin^2r\cos^2r\cos2h}
$$
and $\sin2\theta^*=0$ or
\begin{eqnarray}
\begin{aligned}
&\tan2\theta^*\\ \nonumber
=&\frac{2[-1+\sin^22q\cos^2(g-\alpha^*)+\sin^22r\cos^2(h-\alpha^*)]}{\sin4q\cos(g-\alpha^*)+\sin4r\cos(h-\alpha^*)}.
\end{aligned}
\end{eqnarray}

From (\ref{tm2}) we have
\begin{eqnarray}\label{iib}
\begin{aligned}
&H(\mathcal{A}_1)+H(\mathcal{A}_2)+H(\mathcal{A}_3)\\
=&-p_1\log_2 p_1-p_2\log_2 p_2-q_1\log_2 q_1-q_2\log_2 q_2\\
&-r_1\log_2 r_1-r_2\log_2 r_2\\
\geq &\min\limits_{i,\theta,\alpha}[\mathbf{P}_1^{(i)}(\theta)+\mathbf{Q}_1^{(i)}(\theta,\alpha)+\mathbf{R}_1^{(i)}(\theta,\alpha)]\\
\geq &\min\limits_i L_i,
\end{aligned}
\end{eqnarray}
where if $\min\{\theta_{i-1},\theta_i\}\leq \theta_i^*\leq \max\{\theta_{i-1},\theta_i\},$
\begin{eqnarray}
\begin{aligned}
L_i=&\{\mathbf{P}_1^{(i)}(\theta_i^*)+\mathbf{Q}_1^{(i)}(\theta_i^*,\alpha_i^*)+\mathbf{R}_1^{(i)}(\theta_i^*,\alpha_i^*),\\ \nonumber
&\mathbf{P}_1^{(i)}(\theta_{i})+\mathbf{Q}_1^{(i)}(\theta_{i},\alpha_i^{\star})+\mathbf{R}_1^{(i)}(\theta_{i},\alpha_i^{\star}),\\ \nonumber
&\mathbf{P}_1^{(i)}(\theta_{i-1})+\mathbf{Q}_1^{(i)}(\theta_{i-1},\alpha_{i-1}^{\star})\\ \nonumber
&+\mathbf{R}_1^{(i)}(\theta_{i-1},\alpha_{i-1}^{\star})\},
\end{aligned}
\end{eqnarray}
otherwise,
\begin{eqnarray}
\begin{aligned}
L_i=&\{\mathbf{P}_1^{(i)}(\theta_{i})+[\mathbf{Q}_1^{(i)}(\theta_{i},\alpha_i^{\star})+\mathbf{R}_1^{(i)}(\theta_{i},\alpha_i^{\star})],\\ \nonumber
&\mathbf{P}_1^{(i)}(\theta_{i-1})+[\mathbf{Q}_1^{(i)}(\theta_{i-1},\alpha_{i-1}^{\star})\\ \nonumber
&+\mathbf{R}_1^{(i)}(\theta_{i-1},\alpha_{i-1}^{\star})]\}.
\end{aligned}
\end{eqnarray}
$\mathbf{P}_1^{(i)}(\theta),$ $\mathbf{Q}_1^{(i)}(\theta,\alpha),$ $\mathbf{R}_1^{(i)}(\theta,\alpha)$ and $\theta_i^*$ and  $\alpha_i^*$ are given in Appendix A with $p=1$.
$\alpha_{i-1}^{\star}$ and $\alpha_i^{\star}$ are extreme points of

$\mathbf{P}_1^{(i)}(\theta_{i-1})+\mathbf{Q}_1^{(i)}(\theta_{i-1},\alpha)+\mathbf{R}_1^{(i)}(\theta_{i-1},\alpha)$

and

$\mathbf{P}_1^{(i)}(\theta_{i})+\mathbf{Q}_1^{(i)}(\theta_{i},
\alpha)+\mathbf{R}_1^{(i)}(\theta_{i},\alpha)$, respectively.

\noindent{\textit{Example 1:}} In Ref. \cite{Rudnicki},  the authors consider qubit case with three measurements $\mathcal{A}_k~(k=1,2,3)$  given by vectors
$|a_1^{(1)}\rangle=\{1,0\},$ $|a_2^{(1)}\rangle=\{0,1\}$;  $|a_1^{(2)}\rangle=\{\cos\theta,\sin\theta\},$ $|a_2^{(2)}\rangle=\{\sin\theta, -\cos\theta\}$;
and $|a_1^{(3)}\rangle=\{\cos\theta, i \sin\theta\},$ $|a_2^{(3)}\rangle=\{\sin\theta, -i\cos\theta\}$, respectively.
From (\ref{iia}), we have that $H(\mathcal{A}_1)+H(\mathcal{A}_2)+H(\mathcal{A}_3)$ is lower bounded by
\begin{equation}
\begin{aligned}
\min\{&\frac{1}{4}(7 - \frac{18\cos^32\theta}{\sqrt{8+\cos^22\theta}} - \cos4\theta - \frac{8\sin2\theta\sin4\theta}{\sqrt{8+\cos^22\theta}}),\\ \nonumber
&\frac{1}{4}(7 + \frac{18\cos^32\theta}{\sqrt{8+\cos^22\theta}} - \cos4\theta + \frac{8\sin2\theta\sin4\theta}{\sqrt{8+\cos^22\theta}})\}.\\ \nonumber
\end{aligned}
\end{equation}

From (\ref{iib}), we have also the lower bound by stepwise linear function.
Figure. 2 shows that our bound by quadratic function is better than (\ref{Ru283}) and (\ref{Xiaob}) when $\theta\in(0.5,1.0)$.
Our bound by stepwise linear function is better than the bounds (\ref{Ru283}) and (\ref{Xiaob}) for $0\leq\theta\leq \frac{\pi}{2}.$

\begin{figure}[htb]
\centering
\includegraphics[width=5.5cm]{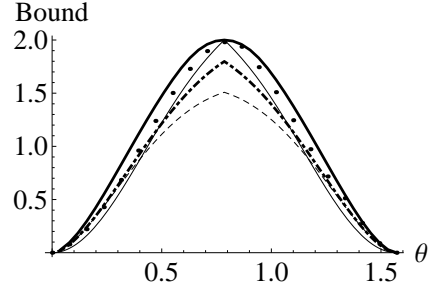}
\caption{Our bound (\ref{iia}) is represented by the solid line, and our bound (\ref{iib}) is plotted by the dotted line with $n=32$. The bound (\ref{Ru283})
from Ref. \cite{Rudnicki} is shown by the dot-dashed line, and the lower bound in (\ref{Xiaob}) is depicted by the dashed line. The thick solid line is the value of $H(\mathcal{A}_1)+H(\mathcal{A}_2)+H(\mathcal{A}_3)$.}
\end{figure}

\noindent{\textit{Case III:}} Qutrit states with three measurements $\mathcal{A}_k$ given by $\{ |a_j^{(k)}\rangle \}~ (k=1,2,3)$.
Denote $\mathbf{p}=\{p_1,p_2,p_3\}$ with
$p_1=\sin^2\theta\cos^2\phi$, $p_2=\sin^2\theta\sin^2\phi$ and $p_3=\cos^2\theta$. Let  $\mathbf{q}=\{q_1,q_2,q_3\}$ and $\mathbf{r}=\{r_1,r_2,r_3\}$, with
\begin{align}\nonumber
q_i&=|\langle a^{(1)}_1|a^{(2)}_i\rangle \sqrt{p_1} +\langle a^{(1)}_2|a^{(2)}_i\rangle \sqrt{p_2} e^{i\alpha}\\ \nonumber
&+\langle a^{(1)}_3|a^{(2)}_i\rangle\sqrt{p_3} e^{i\beta}|^2,\\[2mm] \notag
r_i&=|\langle a^{(1)}_1|a^{(3)}_i\rangle \sqrt{p_1}+\langle a^{(1)}_2|a^{(3)}_i\rangle \sqrt{p_2}  e^{i\alpha}\\ \nonumber
&+\langle a^{(1)}_3|a^{(3)}_i\rangle\sqrt{p_3} e^{i\beta}|^2.\notag
\end{align}
From (\ref{tm1}) we get
\begin{align}\nonumber
&H(\mathcal{A}_1)+H(\mathcal{A}_2)+H(\mathcal{A}_3)\\ \nonumber
\geq & \min\limits_{\theta,\phi,\alpha,\beta}[2(1-\sum\limits_{i=1}^3 p_i^2)+2(1-\sum\limits_{i=1}^3 q_i^2)+2(1-\sum\limits_{i=1}^3 r_i^2)] \\ \nonumber
=& \min\limits_{\theta,\phi,\alpha,\beta} 2[3-\sum\limits_{i=1}^3 (p_i^2+q_i^2+r_i^2)]\\ \nonumber
\geq & \min\{[(3-\sum\limits_{i=1}^3 (p_i^2+q_i^2+r_i^2)]|_{\theta=\theta^*,\phi=\phi^*,\alpha=\alpha^*,\beta=\beta^*}\\ \nonumber
& 2[3-\sum\limits_{i=1}^3 (p_i^2+q_i^2+r_i^2)]|_{\theta=\frac{m_1\pi}{2},\phi=\frac{m_2\pi}{2},\alpha=\frac{m_3\pi}{2},\beta=\frac{m_4\pi}{2}}\},
\end{align}
where
$m_j~(j=1,2,3,4)$ are integers, while $\theta^*$, $\phi^*$, $\alpha^*$, $\beta^*$ are the stationary points of the function $2[3-\sum\limits_{i=1}^3 (p_i^2+q_i^2+r_i^2)]$.

From (\ref{tm2}) we have the following uncertainty relation based on stepwise linear function,
\begin{eqnarray}
\begin{aligned}
& H(\mathcal{A}_1)+H(\mathcal{A}_2)+H(\mathcal{A}_3)\\ \nonumber
\geq &\min\limits_{\theta,\phi,\alpha,\beta,i,j} [\mathbf{P}^{(i,j)}_1(p_1,p_2)+\mathbf{Q}^{(i,j)}_1(q_1,q_2)+\mathbf{R}^{(i,j)}_1(r_1,r_2)]\\ \nonumber
\geq & \min\limits_{i,j}L_{ij},
\end{aligned}
\end{eqnarray}
where
$\mathbf{P}^{(i,j)}_1(p_1,p_2)$, $\mathbf{Q}^{(i,j)}_1(q_1,q_2)$, $\mathbf{R}^{(i,j)}_1(r_1,r_2)$  and $L_{ij}$ are given in Appendix B.

\noindent{\textit{Example 2:}} Consider qutrite states with three measurements $\mathcal{A}_k~(k=1,2,3)$ given by the vectors \cite{Xiao}:
$|a_1^{(1)}\rangle=\{1,0,0\},$ $|a_2^{(1)}\rangle=\{0,1,0\}$, $|a_3^{(1)}\rangle=\{0,0,1\}$; $|a_1^{(2)}\rangle=\{\frac{1}{\sqrt{2}},0,-\frac{1}{\sqrt{2}}\},$ $|a_2^{(2)}\rangle=\{0,1,0\},$ $|a_3^{(2)}\rangle=\{\frac{1}{\sqrt{2}},0,\frac{1}{\sqrt{2}}\}$;
$|a_1^{(3)}\rangle=\{\sqrt{a},e^{i\phi}\sqrt{1-a},0\},$ $|a_2^{(3)}\rangle=\{\sqrt{1-a},-e^{i\phi}\sqrt{a},0\},$ $|a_3^{(3)}\rangle=\{0,0,1\}$ with $\phi=\frac{\pi}{2},$ respectively.
A bound $H_s(\omega_0)$, $\omega_0=\{\Omega_1,1-\Omega_1\}$, is obtained in Ref. \cite{Xiao}. Our bound based on quadratic function for this case is $4 a(1-a)$; see Fig. 3
for comparison.

\begin{figure}[htb]
  \centering
  \includegraphics[width=5.5cm]{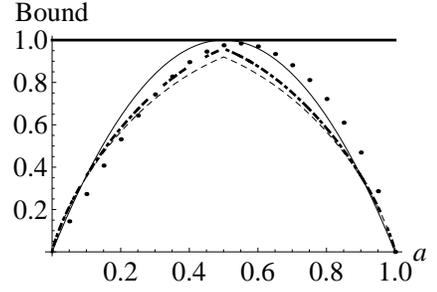}
	\caption{Our bound based on quadratic function (stepwise function) is represented by the solid (dotted line with $n=16$). The bound (\ref{Xiaob}) in Ref. \cite{Xiao} is represented by the dashed line,
and the bound (\ref{Ru283}) in Ref. \cite{Rudnicki} is plotted by the  dot-dashed line. The thick solid line stands for the value of $\sum_{k=1}^3H(\mathcal{A}_k)$.}
\end{figure}
	
\section{Entropic uncertainty relation with state-dependent bound}
In this section we illustrate our state-dependent bounds, showing how our results provide a better estimation for the sum of entropies.
Consider an $N$-dimensional quantum state $\rho$ with  spectral decomposition $\rho = \sum\limits_{i=1}^{d} p^0_i |v_i\rangle\langle v_i|$.
Performing two measurements $\mathcal{A}_1$ and $\mathcal{A}_2$ given by bases $|a^{(1)}_j\rangle$ and $|a^{(2)}_j\rangle$ $(j=1,2,\cdots,d)$, we have the following entropic uncertainty relation:
\begin{equation}\label{11}
H(\mathcal{A}_1)+H(\mathcal{A}_2)\geq \min\limits_{\theta_{ij},\alpha_{ij}} [H(\mathcal{A}_1)+H(\mathcal{A}_2)],
\end{equation}
with the probabilities
\begin{equation}
\begin{aligned}
p_j&=\sum\limits_i p^0_i |\langle v_i|a^{(1)}_j\rangle|^2=\sum\limits_i p^0_i \cos^2\theta_{ij}
\end{aligned}
\end{equation}
and
\begin{equation}
\begin{aligned}
q_j&= \sum\limits_i p^0_i |\langle v_i|a^{(2)}_j\rangle|^2\\ \nonumber
&=\sum\limits_i p^0_i |\langle a^{(1)}_1|a^{(2)}_j\rangle\langle v_i|a^{(1)}_1\rangle+\cdots +\langle a^{(1)}_N|a^{(2)}_j\rangle\langle v_i|a^{(1)}_N\rangle|^2\\ \nonumber
&=\sum\limits_i p^0_i |\sum\limits_{k=1}^d\langle a^{(1)}_k|a^{(2)}_j\rangle\langle v_i|a^{(1)}_k\rangle|^2\\ \nonumber
&=\sum\limits_i p^0_i |\sum\limits_{k=1}^d\langle a^{(1)}_k|a^{(2)}_j\rangle|\cos\theta_{ik}|e^{\mathbf{i}\alpha_{ik}}|^2,\nonumber
\end{aligned}
\end{equation}
respectively.
The above results have been derived based on the fact
that $\langle v_i|a^{(1)}_j\rangle=\cos\theta_{ij}e^{\mathbf{i}\alpha_{ij}}$ and $\sum\limits_j\cos^2\theta_{ij}=\sum\limits_i\cos^2\theta_{ij}=1$.
The minimum in (\ref{11}) runs over all the measurements  $|a_i'\rangle= \mathbf{T}|a_i\rangle$ and $|b_i'\rangle= \mathbf{T}|b_i\rangle$ under base
transformation $T$. In general, we have the following theorem:

\noindent{\textit{Theorem II:}} Given measurements $\mathcal{A}_k$ ($k=1,2,\cdots,N$) with bases $|a^{(k)}_j\rangle$, for arbitrary state $\rho = \sum\limits_{i=1}^{d} p^0_i |v_i\rangle\langle v_i|$,
we have
\begin{equation}\label{tmII}
\sum\limits_{k=1}^N H(\mathcal{A}_k)\geq \min\limits_{\theta_{ij},\alpha_{ij}}\sum\limits_{k=1}^N H(\mathcal{A}_{k})\geq \tilde{H}(\Omega)
\end{equation}
with the corresponding probabilities
\begin{equation}
\begin{aligned}
p^{(1)}_j&= \sum\limits_i p^0_i |\langle v_i|a^{(1)}_j\rangle|^2 =\sum\limits_i p^0_i\cos^2\theta_{ij}
\end{aligned}
\label{eq:1}
\end{equation}
and
\begin{equation}
\begin{aligned}
p^{(k)}_j& = \sum\limits_i p^0_i |\langle v_i|a^{(k)}_j\rangle|^2\\ \nonumber
&=\sum\limits_i p^{0}_i |\sum \limits_{s=1}^N \langle a_s^{(1)}|a_j^{(k)}\rangle\langle v_i|a_s^{(1)}\rangle|^2\\ \nonumber
&=\sum\limits_i p^{0}_i |\sum \limits_{s=1}^N \langle a_s^{(1)}|a_j^{(k)}\rangle|\cos\theta_{is}|e^{\mathbf{i}\alpha_{is}}|^2,
\end{aligned}
\label{eq:2}
\end{equation}
where the probability vectors satisfy the direct-sum majorization uncertainty relation,
$$
\bigoplus\limits_k p^{(k)}\prec \Omega=\{\omega_1, \omega_2 - \omega_1, \omega_3 - \omega_2, \cdots, \omega_d - \omega_{d-1}\},
$$
and $\omega_{i}$ ($i=1, 2, \cdots, d$) are defined by
\begin{equation}
\begin{aligned}
&\omega_1=\max\limits_{j,k}\{p^{(k)}_{j}\},\\ \nonumber
&\omega_2=\max\limits_{j_1,k_1,j_2,k_2}\{p^{(k_1)}_{j_1} + p_{j_2}^{(k_2)}\},\\ \nonumber
&\cdots \\ \nonumber
&\omega_d=d.
\end{aligned}
\end{equation}

Let us consider the following detailed case.

\noindent{\textit{Case IV:}}
Consider qubit states with the spectral decomposition $\rho=p |v\rangle+(1-p)|v_{\bot}\rangle\langle v_{\bot}|$, and
two measurements $\mathcal{A}_k$ given by $|a^{(k)}_j\rangle\langle a^{(k)}_j|$, $j=1,2$, $k=1,2$.
We have
\begin{equation}\label{majm2}
\begin{aligned}
&H(\mathcal{A}_1)+H(\mathcal{A}_2)\\
=&-p_1\log_2 p_1-p_2\log_2p_2-q_1\log_2 q_1-q_2\log_2q_2\\
\geq & -\omega_1\log_2\omega_1-(\omega_2-\omega_1)\log_2(\omega_2-\omega_1)\\
&-(\omega_3-\omega_2)\log_2(\omega_3-\omega_2)-(\omega_4-\omega_3)\log_2(\omega_-\omega_3)\\
=&-(\omega_2-\omega_1)\log_2(\omega_2-\omega_1)-(\omega_3-\omega_2)\log_2(\omega_3-\omega_2)\\
&+ S(\rho),
\end{aligned}
\end{equation}
where $p_1= p \cos^2\theta + (1-p) \sin^2\theta,$ $p_2=p \sin^2\theta + (1-p) \cos^2\theta,$
$q_1= p \cos^2(q-\theta) + (1-p) \sin^2(q-\theta),$ and $q_2=p \sin^2(q-\theta) + (1-p) \cos^2(q-\theta)$,
while $\cos q=|\langle a^{(1)}_1|a^{(2)}_1\rangle|$ and $\sin q=|\langle a^{(1)}_2|a^{(2)}_1\rangle|$,
\begin{equation}
\begin{aligned}
\omega_1=&\max\{p,1-p\},\\ \nonumber
\omega_2=&\max\{1+(2p-1)\cos q, 1+(2p-1)\sin q, \\ \nonumber
&1+(1-2p)\sin q, 1+(1-2p)\cos q\},\\ \nonumber
\omega_3=&\max\{1+p,2-p\},\\ \nonumber
\omega_4=&2. \nonumber
\end{aligned}
\end{equation}

For the case $\rho=p |v\rangle+(1-p)|v_{\bot}\rangle\langle v_{\bot}|$ with three measurements $|a^{(k)}_j\rangle\langle a^{(k)}_j|$, $k=1,2,3$, $j=1,2$,
we have
\begin{equation}\label{majm3}
\begin{aligned}
&H(\mathcal{A}_1)+H(\mathcal{A}_2)+H(\mathcal{A}_3)\\
=&-p_1\log_2 p_1-p_2\log_2 p_2-q_1\log_2 q_1-q_2\log_2 q_2\\
&-r_1\log_2 r_1-r_2\log_2 r_2\\
\geq & -\omega_1\log\omega_1-(\omega_2-\omega_1)\log_2 \omega_2-\omega_1\\
&\cdots -(\omega_5-\omega_4)\log_2 (\omega_5-\omega_4) -(3-\omega_5)\log_2 (3-\omega_5)\\
=& -(\omega_2-\omega_1)\log_2 \omega_2-\omega_1 \cdots -(\omega_5-\omega_4)\log_2 (\omega_5-\omega_4)\\
&+S(\rho),
\end{aligned}
\end{equation}
where $p_i,q_i,r_i~(i=1,2)$ and $w_i~(i=1,2,\cdots,6)$ are listed in Appendix C.







In terms of the stepwise linear function, we have the following conclusion.

\noindent{\textit{Theorem III:}} Given an arbitrary state $\rho$ and measurements $\mathcal{A}_k$ ($k=1,2,\cdots,N$) associated to vectors $|a^{(k)}_j\rangle$, respectively, we have
\begin{equation}\label{tmIII}
\begin{aligned}
\sum\limits_{k=1}^N H(\mathcal{A}_k)\geq &\min\limits_{\theta_{ij},\alpha_{ij}}\sum\limits_{k=1}^N H(p^{(k)})\\
\geq &\min\limits_{\theta_{ij},\alpha_{ij},s}\mathbf{P}^{(s)}_{1,1}(\theta_{ij})+\sum\limits_{k=2}^N \mathbf{P}^{(s)}_{1,k}(\theta_{ij},\alpha_{ij}),
\end{aligned}
\end{equation}
where $\mathbf{P}^{(s)}_{1,1}(\theta_{ij})$ and $\mathbf{P}^{(s)}_{1,k}(\theta_{ij})$ are stepwise linear functions for $\{p_j^{(1)}\}=\{\tr(\rho.|a^{(1)}_j\rangle\langle a^{(1)}_j)|\}$ and $\{p_j^{(k)}\}=\{\tr(\rho.|a^{(k)}_j\rangle\langle a^{(k)}_j)|\}$.

As a particular case, from (\ref{tmIII}) we have the following entropic uncertainty relations.

\noindent{\textit{Case V:}} For two measurement case, one has
\begin{equation}\label{step2}
\begin{aligned}
&H(\mathcal{A}_1)+H(\mathcal{A}_2)\\
=&-p_1\log_2 p_1-p_2\log_2p_2-q_1\log_2 q_1-q_2\log_2q_2\\
\geq & \min\limits_{\theta,i} (\mathbf{P}_1^{(i)}(\theta)+\mathbf{Q}_1^{(i)}(\theta))\\
\geq & \min\limits_i L_i,
\end{aligned}
\end{equation}
where if $\min\{\theta_{i-1},\theta_i\}\leq \theta_i^*\leq \max\{\theta_{i-1},\theta_i\},$
\begin{equation}
\begin{aligned}\\ \nonumber
L_i=&\{\mathbf{P}_1^{(i)}(\theta_i^{*})+\mathbf{Q}_1^{(i)}(\theta_i^{*}), \mathbf{P}_1^{(i)}(\frac{k\pi}{2})+\mathbf{Q}_1^{(i)}(\frac{k\pi}{2}),\\ \nonumber
&\mathbf{P}_1^{(i)}(\theta_{i-1})+\mathbf{Q}_1^{(i)}(\theta_{i-1}), \mathbf{P}_1^{(i)}(\theta_i)+\mathbf{Q}_1^{(i)}(\theta_i)\}
\end{aligned}
\end{equation}
otherwise
\begin{equation}
\begin{aligned}\\ \nonumber
L_i=&\{\mathbf{P}_1^{(i)}(\frac{k\pi}{2})+\mathbf{Q}_1^{(i)}(\frac{k\pi}{2}),\mathbf{P}_1^{(i)}(\theta_{i-1})+\mathbf{Q}_1^{(i)}(\theta_{i-1}), \\ \nonumber &\mathbf{P}_1^{(i)}(\theta_i)+\mathbf{Q}_1^{(i)}(\theta_i)\}
\end{aligned}
\end{equation}
and $\mathbf{P}_1^{(i)}$, $\mathbf{Q}_1^{(i)}$ and $\theta_i^{*}$ are given in Appendix D.

\noindent{\textit{Example 3:}} Consider the qubit state with eigenvalues ${p,1-p}$ and two measurements $\mathcal{A}_k~(k=1,2)$: $|a_j^{(k)}\rangle,~j=1,2$.
We compare our result in this case with the existing bounds in Fig. 4. From Fig. 4 we see that when $p=0.2,$ our bound (\ref{majm2}) is better than  the bound (\ref{korze}) in Ref. \cite{Korzekwa} and the bound (\ref{Ru282}) in
Ref. \cite{Rudnicki}  for $0.8 \leq c\leq 1,$ and is almost the same as the bound (\ref{Yuanb}) in Ref. \cite{Yuan}. Our bound (\ref{step2}) for $n=32$ is better than all the other bounds. From Fig. 5 we see that when $p=0.055,$ our bound (\ref{majm2}) is better than the bound (\ref{korze}) in Ref. \cite{Korzekwa} and  the bound (\ref{Ru282}) in
Ref. \cite{Rudnicki} for $0.78 \leq c\leq 1,$, and is almost the same as the bound (\ref{Yuanb}) in Ref. \cite{Yuan}. Our bound (\ref{step2}) with $n=8$ is better than all the other bounds.

\begin{figure}[htpb]
\centering
\includegraphics[width=5.5cm]{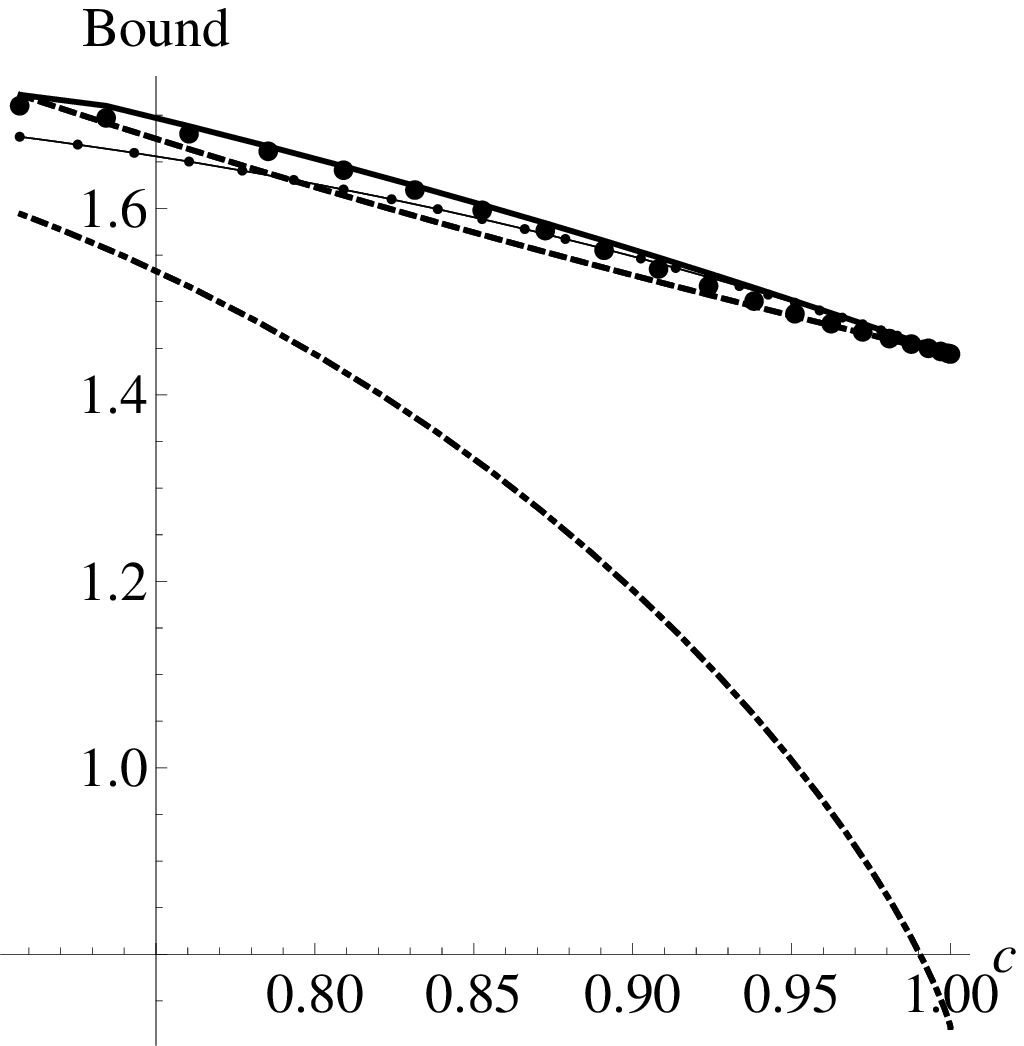}
\caption{Our bound (\ref{majm2}) by majorization is represented by the solid line, and our bound (\ref{step2}) by the stepwise linear function, with $n=32$, is represented by the big size dotted line.
The bound (\ref{Yuanb}) in Ref. \cite{Yuan} is represented by the small size doted line, the bound (\ref{Ru282}) in Ref. \cite{Rudnicki}
is represented by the thick dot-dashed line, the bound (\ref{korze}) in Ref. \cite{Korzekwa} is represented by the thick dashed line, and the thick solid line is for $\sum_{k=1}^2H(\mathcal{A}_k)$ with $p=0.2$.}
\end{figure}

\begin{figure}[htpb]
\centering
\includegraphics[width=5.5cm]{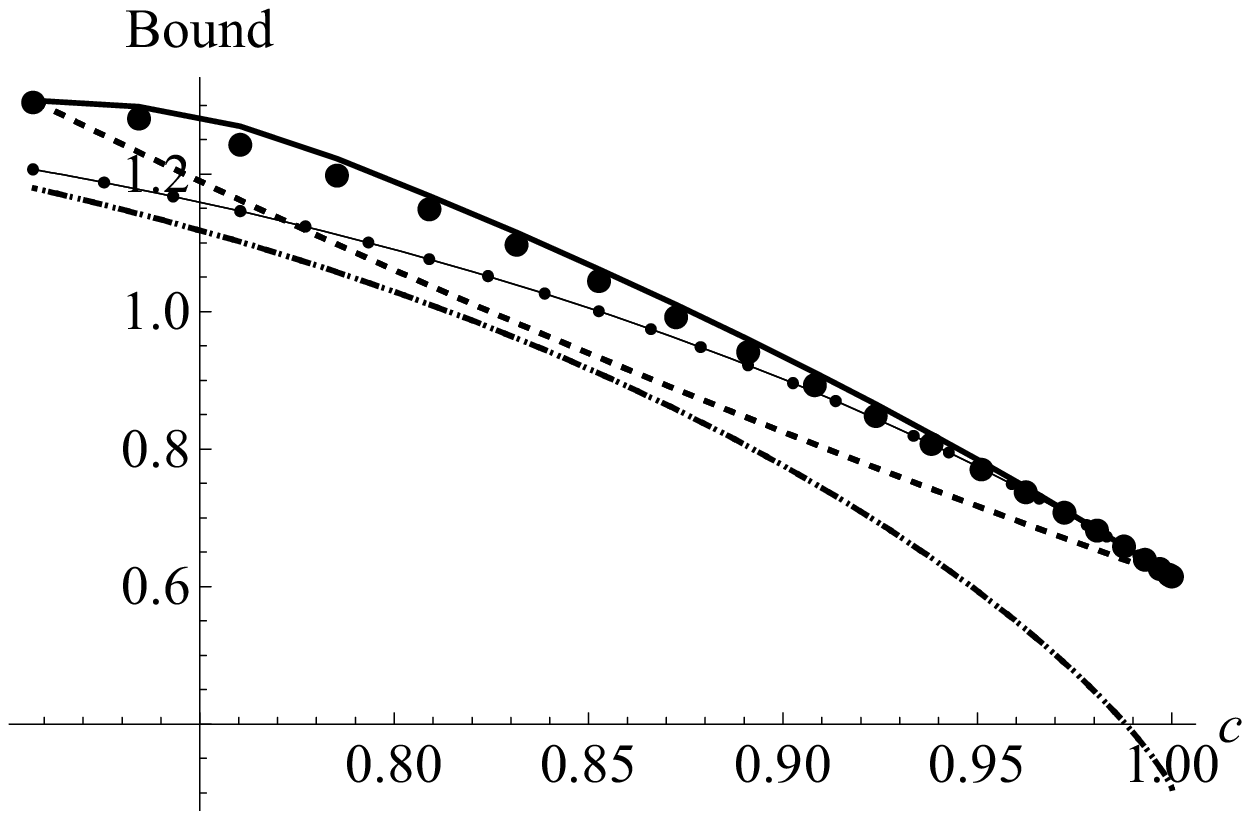}
\caption{Our bound (\ref{majm2}) by majorization techniques is represented by the solid line, and our bound (\ref{step2}) with $n=8$ by the stepwise linear function is represented by the  big size dotted line. The bound (\ref{Yuanb}) in Ref. \cite{Yuan} is represented by the small size dotted line, the bound (\ref{Ru282}) in Ref. \cite{Rudnicki}
is represented by the thick dot-dashed line, the bound (\ref{korze}) in Ref. \cite{Korzekwa} is represented by the  thick dashed line,
and the thick solid line is for $\sum_{k=1}^2H(\mathcal{A}_k)$ with $p=0.055$.}
\end{figure}

\noindent{\textit{Case VI:}} For the case of three measurements one gets
\begin{equation}\label{step3}
\begin{aligned}
&H(\mathcal{A}_1)+H(\mathcal{A}_2)+H(\mathcal{A}_3)\\
=&-p_1\log_2 p_1-p_2\log_2 p_2-q_1\log_2 q_1-q_2\log_2 q_2\\
&-r_1\log_2 r_1-r_2\log_2 r_2\\
\geq &\min\limits_{i,\theta,\alpha}[\mathbf{P}_1^{(i)}(\theta)+\mathbf{Q}_1^{(i)}(\theta,\alpha)+\mathbf{R}_1^{(i)}(\theta,\alpha)]\\
\geq & L_i
\end{aligned}
\end{equation}
if $\min\{\theta_{i-1},\theta_i\}\leq \theta_i^*\leq \max\{\theta_{i-1},\theta_i\},$
\begin{eqnarray}
\begin{aligned}
L_i=&\{\mathbf{P}_1^{(i)}(\theta_i^*)+\mathbf{Q}_1^{(i)}(\theta_i^*,\alpha_i^*)+\mathbf{R}_1^{(i)}(\theta_i^*,\alpha_i^*),\\ \nonumber
&\mathbf{P}_1^{(i)}(\theta_{i})+\min\limits_{\alpha_i}[\mathbf{Q}_1^{(i)}(\theta_{i},\alpha_i)+\mathbf{R}_1^{(i)}(\theta_{i},\alpha_i)],\\ \nonumber
&\mathbf{P}_1^{(i)}(\theta_{i-1})+\min\limits_{\alpha_i}[\mathbf{Q}_1^{(i)}(\theta_{i-1},\alpha_{i-1})\\ \nonumber
&+\mathbf{R}_1^{(i)}(\theta_{i-1},\alpha_{i-1})]\},
\end{aligned}
\end{eqnarray}
otherwise,
\begin{eqnarray}
\begin{aligned}
L_i=&\{\mathbf{P}_1^{(i)}(\theta_{i})+\min\limits_{\alpha_i}[\mathbf{Q}_1^{(i)}(\theta_{i},\alpha_i)+\mathbf{R}_1^{(i)}(\theta_{i},\alpha_i)],\\ \nonumber
&\mathbf{P}_1^{(i)}(\theta_{i-1})+\min\limits_{\alpha_{i-1}}[\mathbf{Q}_1^{(i)}(\theta_{i-1},\alpha_{i-1})\\ \nonumber
&+\mathbf{R}_1^{(i)}(\theta_{i-1},\alpha_{i-1})]\},
\end{aligned}
\end{eqnarray}
where $\mathbf{P}_1^{(i)}(\theta)$, $\mathbf{Q}_1^{(i)}(\theta,\alpha),$ $\mathbf{R}_1^{(i)}(\theta,\alpha)$ and $\theta^*,$ $\alpha^*$ are given in Appendix A.

\noindent{\textit{Example 4:}} Consider
qubit states with eigenvalues ${p,1-p}$, and three measurements $\mathcal{A}_i~(i=1,2,3)$  given by
$|a_1^{(1)}\rangle=\{1,0\},$ $|a_2^{(1)}\rangle=\{0,1\},$  $|a_1^{(2)}\rangle=\{\cos\theta,\sin\theta\},$ $|a_2^{(2)}\rangle=\{\sin\theta, -\cos\theta\},$
$|a_1^{(3)}\rangle=\{\cos\theta, i \sin\theta\},$ $|a_2^{(3)}\rangle=\{\sin\theta, -i\cos\theta\}$. We compare our results with the existing bounds in Fig. 6.
\begin{figure}[htb]
\centering
\includegraphics[width=5.5cm]{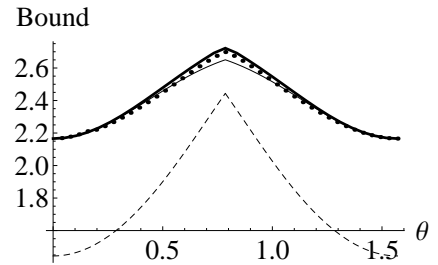}
\caption{Our bound (\ref{majm3}) by majorization is represented by the solid line, and our bound (\ref{step3}) by the stepwise linear function with $n=16$ is represented by
the dotted line. The bound  in Ref. \cite{Xiao} is represented by the dashed line, and
the thick solid line is for $\sum_{k=1}^3H(\mathcal{A}_k)$ with $p=0.2$.}
\end{figure}

When $\{|a^{(k)}_j\rangle\langle a^{(k)}_j|\}$, $k=1,2,3,~j=1,2$, are mutually unbiased bases, straightforward computation shows
\begin{equation}
\begin{aligned}
&\omega_1=\max\{p,1-p\},\\ \nonumber
&\omega_2=\max\{1+\frac{\sqrt{2}}{2}(2p-1),1+\frac{\sqrt{2}}{2}(-2p+1)\},\\ \nonumber
&\omega_3=\max\{\frac{3}{2}+\frac{\sqrt{3}}{2}(2p-1),\frac{3}{2}+\frac{\sqrt{3}}{2}(-2p+1)\},\\ \nonumber
&\omega_4=1+\omega_2,\\ \nonumber
&\omega_5=2+\max\{p,1-p\},\\ \nonumber
&\omega_6=3.
\end{aligned}
\end{equation}

\noindent{\textit{Theorem IV:}} If the three measurements on a qubit state are given by the three mutually unbiased bases, we have
\begin{equation}\label{tmIV}
H(\mathcal{A}_1)+H(\mathcal{A}_2)+H(\mathcal{A}_3) \geq S(\rho)+2.
\end{equation}

{\bf{Proof:}} Assume that the qubit state has eigen-decomposition, $\rho=\sum\limits_{i} p_i |v_i\rangle\langle v_i|$.
where $|v_i\rangle$, $i=1,2$, are orthonomal basis. By choosing the first basis of the three mutually
unbiased bases to be $\mathcal{B}_1=\{|v_1\rangle, |v_2\rangle\}$, one can directly prove (\ref{tmIV}). $\Box$

\section{Discussion and Conclusion}

We have presented improved uncertainty relations based on the Shannon entropy and the majorization techniques.
Analytical formulas are derived for qubits subjected to two or three measurements, which outperform the previous
results. Our results can be also generalized to other forms of uncertainty relation, such as the uncertainty relations of quantum coherence.

Moreover, our improved uncertainty relations can be generalized to the case with the presence of quantum memory.
Let the entropic uncertainty relation without quantum memory be given by
\begin{align}\nonumber
\sum\limits_{k} H(\mathcal{A}_{k}) \geq \mathcal{B}_{ploy},
\end{align}
where $\mathcal{B}_{ploy}$ denotes our lower bounds by polynomial functions.
Now consider bipartite quantum states $\rho_{AB}$ with the subsystem $B$ as the quantum memory.
Generally it is not true that $\sum\limits_{k} H(\mathcal{A}_{k} | B) \geq \mathcal{B}_{ploy} + H(A|B)$,
where, without confusion we still using the symbol $H$ to denote the von Neummann (conditional) entropy.
In order to generalize our bounds to the case with quantum memory, we first define the mutual information $\mathcal{QM}_{k}$ with respect to the $k$-th measurement,
\begin{align}
\mathcal{QM}_{k}=&\sum\limits_{j} \tr(|a_{j}^{(k)}\rangle\langle a_{j}^{(k)}| \rho_{AB})
H(\frac{\tr_{A}(|a_{j}^{(k)}\rangle\langle a_{j}^{(k)}| \rho_{AB})}{\tr(|a_{j}^{(k)}\rangle\langle a_{j}^{(k)}| \rho_{AB})})\notag\\
-&H(\rho_{B}).
\end{align}
$\sum\limits_{k} \mathcal{QM}_{k}$ is a type of quantum correlation measure $\mathcal{Q}_{2}$ \cite{Xiao2017}.
We have the following entropic uncertainty relations in the presence of quantum memory,
\begin{align}
\sum\limits_{k} H(\mathcal{A}_{k} | B) \geq \mathcal{B}_{ploy} + \sum\limits_{k} \mathcal{QM}_{k}.
\end{align}

The uncertainty principle has profound applications in many quantum information processing such  as quantum cryptograph.
The quantum cryptograph in the absence of quantum memory has the possibility for being an eavesdropper to utilizing the quantum correlations.
Using uncertainty relations with the presence of quantum memory one can overcome such eavesdropping.
Our lower bound $\mathcal{B}_{ploy} + \sum\limits_{k} \mathcal{QM}_{k}$ can be used in quantum key distribution directly.
Besides improving the entropic uncertainty relations and extending our results to the case with quantum memory,
our bound is also related to the information exclusion relations \cite{Xiao2016E}.

Moreover, our method can be generalized to the case with Dirac fields: when we consider the bipartite system with Dirac fields,
near the event horizon of a Schwarzschild black hole, the quantity $\sum\limits_{k} \mathcal{QM}_{k}$ provides a better bound
than the previous bounds based on the mutual information. As reported in Ref. \cite{Huang2017}, if the quantum memory
moves away from the black hole, the difference between the total uncertainty and $\mathcal{B}_{ploy} + \sum\limits_{k} \mathcal{QM}_{k}$
remains a constant, independent on the properties of the black hole.

\medskip
{\bf{Acknowledgments}}
This work is supported by the NSFC under Grants No. 11571313, 11371247, 11761141014 and 11675113, and the NSF of Beijing under Grant No. KZ201810028042.

\section{Appendixes}

\subsection{Lower bound by stepwise linear function for mixed states subjected to three measurements}

For any qubit state $\rho = p |v\rangle\langle v| + (1-p)|v^{\perp}\rangle\langle v^{\perp}| $ subjected to three measurements $A_k$ given by ${|a_j^{(k)}\rangle}$, $k=1,2,3$,
set $|\langle v|a_1^{(1)}\rangle|=\cos\theta$. The corresponding probabilities are
\begin{equation}
\begin{aligned}
&p_1=p\cos^2\theta+(1-p)\sin^2\theta,\\ \nonumber
&p_2=p\sin^2\theta+(1-p)\cos^2\theta,\\ \nonumber
&q_1(\theta,\alpha)=p(\cos^2\theta\cos^2 q+\sin^2\theta\sin^2 q+2\sin\theta\cos\theta\sin q\\ \nonumber
&\times\cos q\cos[g-\alpha])+(1-p)(\sin^2\theta\cos^2 q +\cos^2\theta\sin^2 q\\ \nonumber
&-2\sin\theta\cos\theta\sin q\cos q\cos[g-\alpha]),\\ \nonumber
&q_2(\theta,\alpha)=p(\sin^2\theta\cos^2 q+\cos^2\theta\sin^2 q-2\sin\theta\cos\theta\sin q\\ \nonumber
&\times\cos q\cos[g-\alpha])+(1-p)(\cos^2\theta\cos^2 q+\sin^2\theta\sin^2 q\\ \nonumber
&+2\sin\theta\cos\theta\sin q\cos q\cos[g-\alpha]),\\ \nonumber
&r_1(\theta,\alpha)=p(\cos^2\theta\cos^2 r+\sin^2\theta\sin^2 r+2\sin\theta\cos\theta\sin r\\ \nonumber
&\times\cos r\cos[h-\alpha])+(1-p)(\sin^2\theta\cos^2 r+\cos^2\theta\sin^2 r\\ \nonumber
&-2\sin\theta\cos\theta\sin r\cos r\cos[h-\alpha]),\\ \nonumber
&r_2(\alpha)=p(\sin^2\theta\cos^2 r+\cos^2\theta\sin^2 r-2\sin\theta\cos\theta\sin r\\ \nonumber
&\times\cos r\cos[h-\alpha])+(1-p)(\cos^2\theta\cos^2 r+\sin^2\theta\sin^2 r\\ \nonumber
&+2\sin\theta\cos\theta\sin r\cos r\cos[h-\alpha]).
\end{aligned}
\end{equation}
We have
\begin{eqnarray}
\begin{aligned}
\nonumber
&H(\mathcal{A}_1)+H(\mathcal{A}_2)+H(\mathcal{A}_3)\\
=&-p_1\log_2 p_1-p_2\log_2 p_2-q_1\log_2 q_1-q_2\log_2 q_2\\
&-r_1\log_2 r_1-r_2\log_2 r_2\\
\geq &\min\limits_{i,\theta,\alpha} (\mathbf{P}_1^{(i)}(\theta)+\mathbf{Q}_1^{(i)}(\theta,\alpha)+\mathbf{R}_1^{(i)}(\theta,\alpha)),
\end{aligned}
\end{eqnarray}
where $\mathbf{P}_1^{(i)}(\theta)$, $\mathbf{Q}_1^{(i)}(\theta)$ and $\mathbf{R}_1^{(i)}(\theta)$
are given by
\begin{equation}
\begin{aligned}
&\mathbf{P}_1^{(i)}(\theta)=n(l_{i+1}-l_i)(p\cos^2\theta+(1-p)\sin^2\theta-u_i)+l_i,\\ \nonumber
&\mathbf{Q}_1^{(i)}(\theta,\alpha)=\frac{s_{i+1}-s_{i}}{t_{i+1}-t_{i}}(p(\cos^2\theta\cos^2 q+\sin^2\theta\sin^2q\\ \nonumber
&+2\sin\theta\cos\theta\sin q\cos q\cos(g-\alpha))\\ \nonumber
&+(1-p)(\sin^2\theta\cos^2q +\cos^2\theta\sin^2q\\ \nonumber
&-2\sin\theta\cos\theta\sin q\cos q\cos(g-\alpha))-t_{i})+s_{i},\\ \nonumber
&\mathbf{R}_1^{(i)}(\theta,\alpha)=\frac{w_{i+1}-w_{i}}{v_{i+1}-v_{i}}(p(\cos^2\theta\cos^2 r+\sin^2\theta\sin^2r\\ \nonumber
&+2\sin\theta\cos\theta\sin r\cos r\cos(h-\alpha))\\ \nonumber
&+(1-p)(\sin^2\theta\cos^2r+\cos^2\theta\sin^2r\\ \nonumber
&-2\sin\theta\cos\theta\sin r\cos r\cos(h-\alpha))-v_{i})+w_{i},\\ \nonumber
\end{aligned}
\end{equation}
with
\begin{equation}
\begin{aligned}
& u_i=\min\{p,1-p\}+\frac{|1-2p|i}{n},\\ \nonumber
&t_{i}=\max\limits_{\alpha}q_1(\theta_{i},\alpha),v_{i}=\max\limits_{\alpha} r_1(\theta_{i},\alpha),\\ \nonumber
& l_i=-u_i\log_2 u_i-(1-u_i)\log_2(1-u_i),\\ \nonumber
& s_{i}=-t_{i} \log_2 t_{i}-(1-t_{i})\log_2(1-t_{i}),\\ \nonumber
& w_{i}=-v_{i} \log_2 v_{i}-(1-v_{i})\log_2(1-v_{i}),\\ \nonumber
& q=\rm{arccos}|\langle a^{(1)}_1|a_1^{(2)}\rangle|,\\ \nonumber
& r=\rm{arccos}|\langle a^{(1)}_1|a_1^{(3)}\rangle|,\\ \nonumber
& g=\rm{arg}\langle a^{(1)}_1|a_1^{(2)}\rangle-\rm{arg}\langle a^{(1)}_2|a_1^{(2)}\rangle,\\ \nonumber
& h=\rm{arg}\langle a^{(1)}_1|a_1^{(3)}\rangle-\rm{arg}\langle a^{(1)}_2|a_1^{(3)}\rangle,\\ \nonumber
& \theta_i=\arccos[\sqrt{\frac{i}{n}}],  p\geq 1-p, \quad \text{or}, \\ \nonumber
&\theta_i=\arccos[\sqrt{1-\frac{i}{n}}], p\leq 1-p,\\ \nonumber
\end{aligned}
\end{equation}
and the extreme points of  $\mathbf{P}_1^{(i)}(\theta)+\mathbf{Q}_1^{(i)}(\theta,\alpha)+\mathbf{R}_1^{(i)}(\theta,\alpha)$ are given by
\begin{equation}
\begin{aligned}
&\theta_i^{*}=\frac{k\pi}{2},\\ \nonumber
\end{aligned}
\end{equation}
 or
\begin{equation}
\begin{aligned}
&\tan 2\theta_i^{*}\\ \nonumber
&=-\frac{\frac{s_{i+1}-s_{i}}{t_{i+1}-t_{i}}\sin 2q\cos(g-\alpha^*)}{\frac{s_{i+1}-s_{i}}{t_{i+1}-t_{i}}\cos 2q +\frac{w_{i+1}-w_{i}}{v_{i+1}-v_{i}}\cos 2r+n(l_{i+1}-l_i)}\\ \nonumber
&+\frac{\frac{w_{i+1}-w_{i}}{v_{i+1}-v_{i}}\sin 2r\cos(h-\alpha^*)}{\frac{s_{i+1}-s_{i}}{t_{i+1}-t_{i}}\cos 2q +\frac{w_{i+1}-w_{i}}{v_{i+1}-v_{i}}\cos 2r+n(l_{i+1}-l_i)},\\ \nonumber
&\tan\alpha_i^*=\frac{\sin g\sin 2q\frac{s_{i+1}-s_i}{t_{i+1}-t_i}+\sin h\sin 2r\frac{w_{i+1}-i_k}{v_{i+1}-v_i}}
{\cos g \sin 2q\frac{s_{i+1}-s_{i}}{t_{i+1}-t_{i}}+\cos h\sin 2r\frac{w_{i+1}-w_{i}}{v_{i+1}-v_{i}}},
\end{aligned}
\end{equation}
with $\theta^{*}$ satisfying the condition that
$\min\{\theta_{i-1},\theta_{i}\}\leq \theta^{*}\leq \min\{\theta_{i-1},\theta_{i}\}.$

\subsection{Lower bound by stepwise linear function for qutrit pure states subjected to three measurements}
For any qubit state $|\psi\rangle$ subjected to three measurements $A_k$ given by ${|a_j^{(k)}\rangle}$ $k=1,2,3$, the probabilities are
\begin{align}
p_1&=|\langle \psi|a_1^{(1)}\rangle|^2=\sin^2\theta\cos^2\phi\\ \nonumber
p_2&=|\langle \psi|a_2^{(1)}\rangle|^2=\sin^2\theta\sin^2\phi\\ \nonumber
p_3&=|\langle \psi|a_3^{(1)}\rangle|^2=\cos^2\theta\\ \nonumber
q_i&=|\langle a^{(1)}_1|a^{(2)}_i\rangle \sqrt{p_1} +\langle a^{(1)}_2|a^{(2)}_i\rangle \sqrt{p_2} e^{i\alpha}\\ \nonumber
&+\langle a^{(1)}_3|a^{(2)}_i\rangle\sqrt{p_3} e^{i\beta}|^2,\\ \notag
r_i&=|\langle a^{(1)}_1|a^{(3)}_i\rangle \sqrt{p_1}+\langle a^{(1)}_2|a^{(3)}_i\rangle \sqrt{p_2}  e^{i\alpha}\\ \nonumber
&+\langle a^{(1)}_3|a^{(3)}_i\rangle\sqrt{p_3} e^{i\beta}|^2.\notag
\end{align}
The lower bound for stepwise linear function is given by
\begin{align}
&H(\mathcal{A}_1)+H(\mathcal{A}_2)+H(\mathcal{A}_3)\\ \nonumber
\geq &H_s(p_1,p_2)+H_s(q_1,q_2)+H_s(r_1,r_2)\\ \nonumber
\geq &\min\limits_{i,j,\theta,\phi,\alpha,\beta} P_1^{i,j}(p_1,p_2)+Q_1^{i,j}(p_1,p_2)+R_1^{i,j}(p_1,p_2)\\ \nonumber
=&\min\limits_{i,j}L_{i,j}.
\end{align}

If $p^*_1$ and $p^*_2$ satisfy $p^*_1\in h_i, p^*_2\in u_{ij}(p_1)$ or $p^*_2\in v_{ij}(p_1),$ then
\begin{align}
&L_{i,j}\\ \nonumber
=&\min\{P_1^{i,j}(p^*_1,p^*_2)+Q_1^{i,j}(q^*_1,q^*_2)+R_1^{i,j}(r^*_1,r^*_2),\\ \nonumber
&min_1,min_2,min_3,min_4,min_5,min_6\},
\end{align}
where
\begin{align}
&min_1\\ \nonumber
=&\min\limits_{\frac{j-1}{n}\leq p_2\leq \frac{j}{n}}P_1^{i,j}(p_{1x},p_2)+Q_1^{i,j}(p_{1x},p_2)+R_1^{i,j}(p_{1x},p_2),\\ \nonumber
&min_2\\ \nonumber
=&\min\limits_{\frac{i-1}{n}\leq p_1\leq \frac{i}{n}}P_1^{i,j}(p_1,p_{1y})+Q_1^{i,j}(p_1,p_{1y})+R_1^{i,j}(p_1,p_{1y}),\\ \nonumber
&min_3\\ \nonumber
=&\min\limits_{\frac{j-1}{n}\leq p_2\leq\frac{j}{n}}P_1^{i,j}(\frac{i+j-1}{n}-p_2,p_2)\\ \nonumber
+&Q_1^{i,j}(\frac{i+j-1}{n}-p_2,p_2)+R_1^{i,j}(\frac{i+j-1}{n}-p_2,p_2),\\\nonumber
&min_4\\ \nonumber
=&\min\limits_{\frac{j-1}{n}\leq p_2\leq \frac{j}{n}}P_1^{i,j}(p'_{1x},p_2)+Q_1^{i,j}(p'_{1x},p_2)+R_1^{i,j}(p'_{1x},p_2),\\ \nonumber
&min_5\\ \nonumber
=&\min\limits_{\frac{i-1}{n}\leq p_1\leq \frac{i}{n}}P_1^{i,j}(p_1,p'_{2y})+Q_1^{i,j}(p_1,p'_{2y})+R_1^{i,j}(p_1,p'_{2y}),\\ \nonumber
&min_6\\ \nonumber
=&\min\limits_{\frac{j-1}{n}\leq p_2\leq \frac{j}{n}}P_1^{i,j}(\frac{i+j-1}{n}-p_2,p_2)\\ \nonumber
+&Q_1^{i,j}(\frac{i+j-1}{n}-p_2,p_2)+R_1^{i,j}(\frac{i+j-1}{n}-p_2,p_2).
\end{align}

If $p^*_1$ and $p^*_2$ does not satisfy $p^*_1\in h_i, p^*_2\in u_{ij}(p_1)$ and $p^*_2\in v_{ij}(p_1),$ then
\begin{align}
&L_{i,j}\\ \nonumber
=&\min\{min_1,min_2,min_3,min_4,min_5,min_6\}.
\end{align}

The above $min_i$, $1\leq i\leq 6$ can be obtained by using the following extreme points,
\begin{align}
p^*_1&=|\langle \psi|a_1^{(1)}\rangle|^2=\sin^2\theta^*\cos^2\phi^*,\\ \nonumber
p^*_2&=|\langle \psi|a_2^{(1)}\rangle|^2=\sin^2\theta^*\sin^2\phi^*,\\ \nonumber
p^*_3&=|\langle \psi|a_3^{(1)}\rangle|^2=\cos^2\theta^*,\\ \nonumber
q^*_i&=|\langle a^{(1)}_1|a^{(2)}_i\rangle \sqrt{p^*_1} +\langle a^{(1)}_2|a^{(2)}_i\rangle \sqrt{p^*_2} e^{i\alpha^*}\\ \nonumber
&+\langle a^{(1)}_3|a^{(2)}_i\rangle\sqrt{p^*_3} e^{i\beta^*}|^2,\\ \notag
r_i&=|\langle a^{(1)}_1|a^{(3)}_i\rangle \sqrt{p^*_1}+\langle a^{(1)}_2|a^{(3)}_i\rangle \sqrt{p^*_2}  e^{i\alpha^*}\\ \nonumber
&+\langle a^{(1)}_3|a^{(3)}_i\rangle\sqrt{p^*_3} e^{i\beta^*}|^2,\notag
\end{align}
where $\theta^*,$ $\phi^*,$ $\alpha^*$ and $\beta^*$ are the stationary points of $P_1^{i,j}(p_1,p_2)+Q_1^{i,j}(q_1,q_2)+R_1^{i,j}(r_1,r_2)$, and
\begin{eqnarray}
\begin{aligned}
&\mathbf{P}^{(i,j)}_1(p_1,p_2) \\ \nonumber
=&\left\{
\begin{aligned}
&a^{(1)}_{i,j} p_1 + b^{(1)}_{i,j} p_2 + c^{(1)}_{i,j},~ p_1\in h_i,~ p_2 \in u_{ij}(p_1) , \\
&a^{(1)'}_{i,j} p_1 + b^{(1)'}_{i,j} p_2 + c^{(1)'}_{i,j},~ p_1\in h_i,~ p_2 \in v_{ij}(p_1),
\end{aligned}
\right.\\[2mm] \nonumber
&\mathbf{Q}^{(i,j)}_1(q_1,q_2) \\ \nonumber
=&\left\{
\begin{aligned}
&a^{(2)}_{i,j} q_1 + b^{(2)}_{i,j} q_2 + c^{(2)}_{i,j},~ p_1\in h_i,~ p_2 \in u_{ij}(p_1) , \\
&a^{(2)'}_{i,j} q_1 + b^{(2)'}_{i,j} q_2 + c^{(2)'}_{i,j},~ p_1\in h_i,~ p_2 \in v_{ij}(p_1),
\end{aligned}
\right.\\[2mm] \nonumber
&\mathbf{R}^{(i,j)}_1(r_1,r_2) \\ \nonumber
=&\left\{
\begin{aligned}
&a^{(3)}_{i,j} r_1 + b^{(3)}_{i,j} r_2 + c^{(3)}_{i,j},~ p_1\in h_i,~ p_2 \in u_{ij}(p_1) , \\
&a^{(3)'}_{i,j} r_1 + b^{(3)'}_{i,j} r_2 + c^{(3)'}_{i,j},~ p_1\in h_i,~ p_2 \in v_{ij}(p_1).
\end{aligned}
\right. \nonumber
\end{aligned}
\end{eqnarray}
The coefficients $a^{(k)}_{i,j}$, $a^{(k)'}_{i,j},$ $b^{(k)}_{i,j}, b^{(k)'}_{i,j}$ and
$c^{(k)}_{i,j}, c^{(k)'}_{i,j}$ are given by
\begin{eqnarray}
\begin{aligned}
\nonumber
&[a^{(1)}_{i,j},b^{(1)}_{i,j},c^{(1)}_{i,j}]^T=\mathcal{D}_p^{-1}.\mathcal{H}_p^T,
[a^{(1)'}_{i,j},b^{(1)'}_{i,j},c^{(1)'}_{i,j}]^T=\mathcal{D}_p^{'-1}.\mathcal{H}_p^{T'},\\ \nonumber
&[a^{(2)}_{i,j},b^{(2)}_{i,j},c^{(2)}_{i,j}]^T=\mathcal{D}_{q}^{-1}.\mathcal{H}_{q}^T, [a^{(2)'}_{i,j},b^{(2)'}_{i,j},c^{(2)'}_{i,j}]^T=\mathcal{D}_{q}^{'-1}.\mathcal{H}_{q}^{T'},\\ \nonumber
&[a^{(3)}_{i,j},b^{(3)}_{i,j},c^{(3)}_{i,j}]^T=\mathcal{D}_{r}^{-1}.\mathcal{H}_{r}^T, [a^{(3)'}_{i,j},b^{(3)'}_{i,j},c^{(3)'}_{i,j}]^T=\mathcal{D}_{r}^{'-1}.\mathcal{H}_{q}^{r'},\\ \nonumber
\end{aligned}
\end{eqnarray}
where
\begin{eqnarray}
\begin{aligned}
&\mathcal{D}_p=D[p_{1x},p_{1y},p_{2x},p_{2y},p_{3x},p_{3y}],\\ \nonumber
&\mathcal{H}_p=[H(p_{1x},p_{1y}),H(p_{2x},p_{2y}),H(p_{3x},p_{3y})],\\ \nonumber
&\mathcal{D'}_p=D[p'_{1x},p'_{1y},p'_{2x},p'_{2y},p'_{3x},p'_{3y}],\\ \nonumber
&\mathcal{H'}_p=[H(p'_{1x},p'_{1y}),H(p'_{2x},p'_{2y}),H(p'_{3x},p'_{3y})],\\ \nonumber
&\mathcal{D}_q=D[q_{1x},q_{1y},q_{2x},q_{2y},q_{3x},q_{3y}],\\ \nonumber
&\mathcal{H}_q=[H(q_{1x},q_{1y}),H(q_{2x},q_{2y}),H(q_{3x},q_{3y})],\\ \nonumber
&\mathcal{D'}_q=D[q'_{1x},q'_{1y},q'_{2x},q'_{2y},q'_{3x},q'_{3y}],\\ \nonumber
&\mathcal{H'}_q=[H(q'_{1x},q'_{1y}),H(q'_{2x},q'_{2y}),H(q'_{3x},q'_{3y})],\\ \nonumber
&\mathcal{D}_r=D[r_{1x},r_{1y},r_{2x},r_{2y},r_{3x},r_{3y}],\\ \nonumber
&\mathcal{H}_r=[H(r_{1x},r_{1y}),H(r_{2x},r_{2y}),H(r_{3x},r_{3y})],\\ \nonumber
&\mathcal{D'}_r=D[r'_{1x},r'_{1y},r'_{2x},r'_{2y},r'_{3x},r'_{3y}],\\ \nonumber
&\mathcal{H'}_r=[H(r'_{1x},r'_{1y}),H(r'_{2x},r'_{2y}),H(r'_{3x},r'_{3y})]\\ \nonumber
\end{aligned}
\end{eqnarray}
and
\begin{eqnarray}
\begin{aligned}
&p_{1x}=\frac{i-1}{n},~ p_{1y}=\frac{j-1}{n},~ p_{2x}=\frac{i}{n},~ p_{2y}=\frac{j-1}{n},\\ \nonumber
&p_{3x}=\frac{i-1}{n}, ~p_{3y}=\frac{j}{n}, ~p'_{1x}=\frac{i}{n},~ p'_{1y}=\frac{j-1}{n},\\ \nonumber
&p'_{2x}=\frac{i-1}{n}, ~p'_{2y}=\frac{j}{n},~ p'_{3x}=\frac{i}{n},~ p'_{3y}=\frac{j}{n},\\ \nonumber
&q_{ix}=\max\limits_{\alpha,\beta} q_1(p_{ix},p_{iy}),\\ \nonumber
&q_{iy}=\max\limits_{\alpha,\beta} (q_1(p_{ix},p_{iy})+q_2(p_{ix},p_{iy}))-q_{ix}, \\ \nonumber
&q'_{ix}=\max\limits_{\alpha,\beta} q_1(p'_{ix},p'_{iy}),\\ \nonumber
&q'_{iy}=\max\limits_{\alpha,\beta} (q_1(p'_{ix},p'_{iy})+q_2(p'_{ix},p'_{iy}))-q'_{ix}, \\ \nonumber
&r_{ix}=\max\limits_{\alpha,\beta} r_1(p_{ix},p_{iy}), \\ \nonumber
&r_{iy}=\max\limits_{\alpha,\beta} (r_1(p_{ix},p_{iy}) + r_2(p_{ix},p_{iy}))- r_{ix}, \\ \nonumber
&r'_{ix}=\min\limits_{\alpha,\beta} r_1(p'_{ix},p'_{iy}), \\ \nonumber
&r'_{iy}=\max\limits_{\alpha,\beta} (r_1(p'_{ix},p'_{iy}) + r_2(p'_{ix},p'_{iy}))-r'_{ix}, \\ \nonumber
\end{aligned}
\end{eqnarray}
with $h_i=(\frac{i-1}{n}$, $\frac{i}{n}),$ $u_{ij}(x)=(\frac{j-1}{n}$, $\frac{i+j-1}{n} - x)$, $v_{ij}(x)=( \frac{i+j-1}{n} - x,\frac{j}{n}).$

\subsection{Lower bound by majorization techniques for mixed states subjected to three measurements}

For the state $\rho=p |v\rangle+(1-p)|v_{\bot}\rangle\langle v_{\bot}|$ subjected to three measurements $|a^{(k)}_j\rangle\langle a^{(k)}_j|,~k=1,2,3,~j=1,2$,
we have the probabilities $p_i$, $q_i$, $r_i~i=1,2$,
\begin{equation}
\begin{aligned}
p_1=& p\cos^2\theta+(1-p)\sin^2\theta,\\ \nonumber
p_2=&p\sin^2\theta+(1-p)\cos^2\theta,\\ \nonumber
q_1=&p(\cos^2\theta\cos^2 q+\sin^2\theta\sin^2 q+2\sin\theta\cos\theta\sin q\cos q\\ \nonumber
&*\cos[g-\alpha])+(1-p)(\sin^2\theta\cos^2 q +\cos^2\theta\sin^2 q\\ \nonumber
&-2\sin\theta\cos\theta\sin q\cos q\cos[g-\alpha]),\\ \nonumber
q_2=&p(\sin^2\theta\cos^2 q+\cos^2\theta\sin^2 q-2\sin\theta\cos\theta\sin q\cos q\\ \nonumber
&*\cos[g-\alpha])+(1-p)(\cos^2\theta\cos^2 q+\sin^2\theta\sin^2 q\\ \nonumber
&+2\sin\theta\cos\theta\sin q\cos q\cos[g-\alpha]),\\ \nonumber
r_1=&p(\cos^2\theta\cos^2 r+\sin^2\theta\sin^2 r+2\sin\theta\cos\theta\sin r\cos r\\ \nonumber
&*\cos[h-\alpha])+(1-p)(\sin^2\theta\cos^2 r+\cos^2\theta\sin^2 r\\ \nonumber
&-2\sin\theta\cos\theta\sin r\cos r\cos[h-\alpha]),\\ \nonumber
r_2=&p(\sin^2\theta\cos^2 r+\cos^2\theta\sin^2 r-2\sin\theta\cos\theta\sin r\cos r\\ \nonumber
&*\cos[h-\alpha])+(1-p)(\cos^2\theta\cos^2 r+\sin^2\theta\sin^2 r\\ \nonumber
&+2\sin\theta\cos\theta\sin r\cos r\cos[h-\alpha]),
\end{aligned}
\end{equation}
with
$q=\rm{arccos}|\langle a^{(1)}_1|a_1^{(2)}\rangle|,$
$r=\rm{arccos}|\langle a^{(1)}_1|a_1^{(3)}\rangle|,$ $g=\rm{arg}\langle a^{(1)}_1|a_1^{(2)}\rangle-\rm{arg}\langle a^{(1)}_2|a_1^{(2)}\rangle,$
$h=\rm{arg}\langle a^{(1)}_1|a_1^{(3)}\rangle-\rm{arg}\langle a^{(1)}_2|a_1^{(3)}\rangle$.

Therefore,
\begin{equation}
\begin{aligned}
&\omega_1=\max\limits_{\theta,\alpha,i,j,k}\max\{p_i,q_j,r_k\}=\max\{p,1-p\};\\ \nonumber
&\omega_2=\max\limits_{\theta,\alpha,i,j,k,i',j',k'}\max\{1,p_i+q_j,p_i'+r_k,q_j'+r_k'\}\\ \nonumber
&=\max\{1+(2p-1)\cos q, 1+(2p-1)\sin q, \\ \nonumber
&1+(1-2p)\sin q, 1+(1-2p)\cos q, 1+(2p-1)\cos r,\\ \nonumber
&1+(2p-1)\sin r, 1+(1-2p)\sin r, 1+(1-2p)\cos r, \\ \nonumber
&p(\cos^2q+\cos^2r)+(1-p)(\sin^2q+\sin^2r),\\ \nonumber
&p(\sin^2q+\sin^2r)+(1-p)(\cos^2q+\cos^2r),\\ \nonumber
&p(\sin^2q+\cos^2r)+(1-p)(\cos^2q+\sin^2r),\\ \nonumber
&p(\cos^2q+\sin^2r)+(1-p)(\sin^2q+\cos^2r),\\ \nonumber
&(q_1+r_1)_{\theta=\theta_1,\alpha=\alpha_1}, (q_1+r_2)_{\theta=\theta_2,\alpha=\alpha_2}, \\ \nonumber
&(q_2+r_1)_{\theta=\theta_2,\alpha=\alpha_2}, (q_2+r_2)_{\theta=\theta_1,\alpha=\alpha_1}
\}
\end{aligned}
\end{equation}
with
\begin{equation}
\begin{aligned}
&\tan\alpha_1=\frac{\sin2q\sin g+\sin 2r\sin h}{\sin2q\cos g+\sin 2r\cos h},\\ \nonumber
&\tan2\theta_1=\frac{\sin 2q\cos (g-\alpha_1)+\sin 2r\cos (h-\alpha_1)}{\cos 2q+\cos 2r},\\ \nonumber
&\tan\alpha_2=\frac{\sin2q\sin g-\sin 2r\sin h}{\sin2q\cos g-\sin 2r\cos h},\\ \nonumber
&\tan2\theta_2=\frac{\sin 2q\cos (g-\alpha_2)-\sin 2r\cos (h-\alpha_2)}{\cos 2q-\cos 2r};\\ \nonumber
\end{aligned}
\end{equation}
\begin{equation}
\begin{aligned}
&\omega_3\\ \nonumber
=&\max\limits_{\theta,\alpha,i,j,k}\max\{1+p_i,1+q_j,1+r_k\}\\ \nonumber
=&\max\{p(1+\cos^2q+\cos^2r)+(1-p)(\sin^2q+\sin^2r),\\ \nonumber
& p(\sin^2q+\sin^2r)+(1-p)(1+\cos^2q+\cos^2r), \\ \nonumber
&p(1+\cos^2q+\sin^2r)+(1-p)(\sin^2q+\cos^2r), \\ \nonumber
& p(\sin^2q+\cos^2r)+(1-p)(1+\cos^2q+\sin^2r),\\ \nonumber
&p(1+\sin^2q+\cos^2r)+(1-p)(\cos^2q+\sin^2r), \\ \nonumber
& p(\cos^2q+\sin^2r)+(1-p)(1+\sin^2q+\cos^2r),\\ \nonumber
&p(1+\sin^2q+\sin^2r)+(1-p)(\cos^2q+\cos^2r), \\ \nonumber
& p(\cos^2q+\cos^2r)+(1-p)(1+\sin^2q+\sin^2r),\\ \nonumber
&(p_1+q_1+r_1)_{\theta=\theta_1',\alpha=\alpha_1},(p_1+q_1+r_2)_{\theta=\theta_2',\alpha=\alpha_2}\\ \nonumber
&(p_1+q_2+r_1)_{\theta=\theta_3',\alpha=\alpha_2},(p_1+q_2+r_2)_{\theta=\theta_4',\alpha=\alpha_1}\\ \nonumber
&(p_2+q_1+r_1)_{\theta=\theta_4',\alpha=\alpha_1},(p_2+q_1+r_2)_{\theta=\theta_3',\alpha=\alpha_2}\\ \nonumber
&(p_2+q_2+r_1)_{\theta=\theta_2',\alpha=\alpha_2},(p_2+q_2+r_2)_{\theta=\theta_1',\alpha=\alpha_1}\\ \nonumber
&1+\max\{p,1-p\}\}
\end{aligned}
\end{equation}
with
\begin{equation}
\begin{aligned}
&\tan2\theta_1'=\frac{\sin 2q\cos (g-\alpha_1)+\sin 2r\cos (h-\alpha_1)}{\cos\alpha_1(\cos 2q+\cos 2r+1)}\\ \nonumber
&\tan2\theta_2'=\frac{\sin 2q\cos (g-\alpha_2)-\sin 2r\cos (h-\alpha_2)}{\cos\alpha_2(\cos 2q-\cos 2r+1)}\\ \nonumber
&\tan2\theta_3'=\frac{\sin 2q\cos (g-\alpha_2)-\sin 2r\cos (h-\alpha_2)}{\cos\alpha_2(\cos 2q-\cos 2r-1)}\\ \nonumber
&\tan2\theta_4'=\frac{\sin 2q\cos (g-\alpha_1)+\sin 2r\cos (h-\alpha_1)}{\cos\alpha_1(\cos 2q+\cos 2r-1)};\\ \nonumber
\end{aligned}
\end{equation}
\begin{equation}
\begin{aligned}
&\omega_4=1+\omega_2;\\ \nonumber
&\omega_5=2+\max\{p,1-p\};\\ \nonumber
&\omega_6=3.
\end{aligned}
\end{equation}

\subsection{Lower bound by stepwise linear function for mixed states subjected to two measurements}
For qubit states with the spectral decomposition $\rho=p |v\rangle+(1-p)|v_{\bot}\rangle\langle v_{\bot}|$, and
two measurements $\mathcal{A}_k$ given by $|a^{(k)}_j\rangle\langle a^{(k)}_j|$, $j=1,2$, $k=1,2$,
the probabilities are
$p_1= p \cos^2\theta + (1-p) \sin^2\theta,$ $p_2=p \sin^2\theta + (1-p) \cos^2\theta,$
$q_1= p \cos^2(q-\theta) + (1-p) \sin^2(q-\theta),$ and $q_2=p \sin^2(q-\theta) + (1-p) \cos^2(q-\theta)$,
while $\cos q=|\langle a^{(1)}_1|a^{(2)}_1\rangle|$ and $\sin q=|\langle a^{(1)}_2|a^{(2)}_1\rangle|$.
We have the expressions of $\mathbf{P}_1^{(i)},~ \mathbf{Q}_1^{(i)}$ by stepwise linear function and $\theta^{*(i)}$
as follows:
\begin{equation}
\begin{aligned}
&\mathbf{P}_1^{{i}}(\theta)=n(l_{i+1}-l_i)(p\cos^2(\theta)+(1-p)\sin^2(\theta)-u_i)+l_i,\\ \nonumber
&\mathbf{Q}_1^{(i)}(\theta)=\frac{s_{i+1}-s_i}{t_{i+1}-t_i}(p\cos^2(\theta-\gamma)+(1-p)\sin^2(\theta-\gamma)\\ \nonumber
&\hspace{1.5cm}-t_i)+s_i,\\ \nonumber
\end{aligned}
\end{equation}
and
\begin{equation}
\begin{aligned}
&\theta_i=\rm{arccos}\sqrt{1-\frac{i}{n}}~~\rm{for} ~p\leq 1-p,\\ \nonumber
&\rm{and}\\ \nonumber
&\theta_i=\rm{arccos}\sqrt{\frac{i}{n}}~~\rm{for}~p\geq 1-p,\\ \nonumber
& t_i = p\cos^2(\theta_i-\gamma)+(1-p)\sin^2(\theta_i-\gamma),\\ \nonumber
& s_i = -t_i\log_2 t_i -(1-t_i)\log_2(1-t_i),\\ \nonumber
& u_i = \min\{p,1-p\}+\frac{|1-2p|i}{n},\\ \nonumber
& l_i = - u_i\log_2 u_i-(1-u_i)\log_2(1-u_i),\\ \nonumber
& \tan2\theta^{*}_i=\frac{(s_{i+1}-s_i)\sin 2q}{(s_{i+1}-s_i)\cos 2q + n(l_{i+1}-l_i)(t_{i+1}-t_i)},\\ \nonumber
\end{aligned}
\end{equation}
with $\theta^{*}_i$ satisfying the following conditions:
$
\min\{\theta_{i-1},\theta_{i}\}\leq \theta^{*}_i\leq \max\{\theta_{i-1},\theta_{i}\}.
$

\end{document}